\documentclass{appolb}
\usepackage{epsfig}
\usepackage{amssymb}

\newcommand{\ttl}{\f{\tau^2 \th^2}{L^2}}
\newcommand{\alef}{\al'}
\newcommand{\Tau}{{\cal T}}
\newcommand{\vb}{\bar{v}}

\newcommand{\tr}{\mbox{\rm tr }}
\newcommand{\gm}{\gamma}
\newcommand{\xd}{\dot{X}}
\renewcommand{\AA}{{\cal A}}
\newcommand{\DD}{{\cal D}}
\newcommand{\lra}{\longrightarrow}
\newcommand{\xpr}{x_\perp}
\newcommand{\Ttl}{\f{T^2 \th^2}{L^2}}
\renewcommand{\th}{\theta}
\newcommand{\sg}{\sigma}
\newcommand{\numero}[1]{\noindent #1.~}
\newcommand{\la}{\lambda}
\newcommand{\be}{\beta}
\newcommand{\qb}{\bar q}
\newcommand{\bc}{\begin{center}}
\newcommand{\pa}{\partial}
\newcommand{\f}{\frac}
\newcommand{\eqn}{\begin{eqnarray}}
\newcommand{\eqnx}{\end{eqnarray}}
\newcommand{\beg}{\begin{equation}}
\newcommand{\BL}{\begin{Large}}
\newcommand{\EL}{\end{Large}}
\newcommand{\bi}{\begin{itemize}}
\newcommand{\ec}{\end{center}}
\newcommand{\ee}{\end{equation}}
\newcommand{\ei}{\end{itemize}}
\newcommand{\ii}{\bibitem}
\newcommand{\g}{\gamma}

\newcommand{\Ga}{\Gamma}
\newcommand{\de}{\delta}
\newcommand{\om}{\omega}

\newcommand{\nn}{{\cal N}}
\newcommand{\YY}{{\YY}}
\newcommand{\qqqq}{\quad\quad\quad\quad}
\newcommand{\zt}{\tilde{z}}
\newcommand{\dl}{\delta}
\newcommand{\Dl}{\Delta}
\newcommand{\cor}[1]{\left\langle{#1}\right\rangle}
\newcommand{\rsq}{{\mathfrak R^2}}

\newcommand{\alp}{\alpha '}
\newcommand{\n}{\nonumber \\}

\def\al{\alpha}
\def\kb{\bar k^2}
\def\ka{\kappa}

\begin{document}
\title{Introduction to String Theory and Gauge/Gravity duality \\ \\ for 
students in  
QCD and QGP phenomenology\thanks{Presented at the School on QCD, Low-x Physics 
and 
Diffraction, \\
Copanello, Calabria, Italy, July 
2007.}%
\author{Robi Peschanski \thanks{email:{\rm robi.peschanski@cea.fr}}}
\address{Institut de Physique Th{\'e}orique; \\
URA 2306, unit\'e de recherche associ\'ee au CNRS,\\ 
 IPhT, CEA/Saclay,
  91191 Gif-sur-Yvette Cedex, France}
}
\maketitle
\begin{abstract}
String theory has been initially derived from  motivations coming from  strong 
interaction phenomenology, 
but its application faced deep conceptual and practical difficulties. The strong 
interactions  
found their theoretical fundation elsewhere, namely on QCD, the quantum gauge 
field theory of quarks and gluons. Recently, the Gauge/Gravity correspondence 
allowed to initiate a reformulation of  the connection between strings and 
gauge 
field theories, avoiding some of the initial drawbacks and opening the way to 
new insights on the gauge theory at strong coupling and eventually QCD. Among 
others, the recent applications of the Gauge/Gravity correspondence to the 
formation of the QGP, the quark-gluon plasma, in heavy-ion reactions seem to 
provide a physically interesting insight on  phenomenological  features of the 
reactions. In these lectures we will give a simplified introduction to those 
aspects of string theory which, at the origin and in the recent developments, 
are connected to strong interactions, for those students which are starting to 
learn QCD and QGP physics from an experimental or phenomenological 
point of view.
\end{abstract}
  
\eject




\section*{CONTENTS}

\section*{Lecture I: String Theory $via$ Strong Interactions}

\vspace{.1cm}  {\bf \numero{1}}{The origin of String Theory}

\vspace{.1cm}  {\bf \numero{2}}{The Veneziano Formula and Dual Resonance Models} 

\vspace{.1cm}  {\bf \numero{3}}{From Dual Amplitudes to Strings}

\vspace{.1cm}  {\bf \numero{4}}{String Symmetries}

\vspace{.1cm}  {\bf \numero{5}}{Problem: Why 26 (or 10) Dimensions?}

\section*{Lecture II: Gauge/Gravity correspondence}

\vspace{.1cm}  {\bf \numero{6}}{An Open-Closed String Connection}

\vspace{.1cm}  {\bf \numero{7}}{the AdS/CFT Correspondence}

\vspace{.1cm}  {\bf \numero{8}}{Wilson loops, Minimal Surfaces and Confinement}

\vspace{.1cm}  {\bf \numero{9}}{Application: A Dual Model for Dipole Amplitudes}

\section*{Lecture III: Quark Gluon Plasma/Black Hole Duality}

\vspace{.1cm}  {\bf \numero{10}}{QGP Formation and Hydrodynamics}

\vspace{.1cm}  {\bf \numero{11}}{AdS/CFT and  Holographic Hydrodynamics}

\vspace{.1cm}  {\bf \numero{12}}{QGP and Black Holes: From Statics to Dynamics}

\vspace{.1cm}  {\bf \numero{13}}{Thermalization and Isotropization}

\vspace{.1cm}  {\bf \numero{14}}{Outlook}



\eject

\section*{Lecture I: String Theory $via$ Strong Interactions}

\vspace{.2cm}  {\bf \numero{1}} {\it The origin of String Theory}

\vspace{.2cm} 
There is an intimate but rather controversial relationship between strong 
interactions and string theory. As well-known, the birth of string theory 
comes 
from the observation of many puzzling features of strong interaction scattering 
amplitudes from the phenomenological point of view.  In a modern language, we 
call them ``soft'' 
reactions since they involve small-$p_T$ hadrons, and thus a strong coupling 
constant $\alpha_s(p_T)= {\cal O}(1)$ or more
preventing one from using known perturbative techniques of field theory.

It took more or less six years, from 1968 to 1974 starting from the formulation 
of the 
Veneziano amplitude,  to obtain a first consistent formulation of the 
underlying 
string theoretical framework. Strangely enough, it is at the very same time, in 
1974, 
that Quantum Field Theory in the form of Quantum Chromodynamics (QCD), started 
to  be identified as the correct microscopic theoretical fundation of strong 
interactions in terms of quarks and gluons. In fact, it has already been 
realized that the construction of string theory in the physical 
$3\!+\!1$-dimensional Minkowski space has led to numerous difficulties and 
inconsistencies with the observed features of strong interactions. 

It is well-known that starting from that period, string theory and QCD studies 
followed divergent paths, the former being promoted after 1983 
to a serious candidate for the unification of fundamental forces and 
quantum 
gravity and the second showing more and more ability to describe the 
features of quark and gluon interactions at high energy with unprecedenting 
accuracy.

Now, the divorce could have been complete and definitive, when in 1997 appeared 
a new historical twist with the conjecture named ``AdS/CFT correspondence'' and 
its various generalizations and developments involving a new duality relation 
between gauge theories and gravitational interactions in an higher-dimensional 
space. Interestingly, some of the major drawbacks found previously for 
applying string theory to strongly interacting gauge fields have been avoided 
and a new formulation of gauge field theory at strong coupling emerged. Since 
1997, 
the 
developments of the Gauge/Gravity correspondence are numerous.

Many of these new developments are not directly connected to QCD, which indeed 
does not admit for the moment a correct dual formulation. However, they open the 
way 
for new tools for computing amplitudes and other observables of  gauge field 
theories 
in terms of their 
gravity dual. One very promising aspect of this connection 
concerns the 
formation of a Quark-Gluon Plasma (QGP) in  heavy-ion reactions. Indeed, the 
phenomenological features coming from the experiments at RHIC point to the  
formation of a strongly coupled plasma of deconfined gauge fields. In this case, 
one may expect that features of the AdS/CFT correspondence may be relevant. 
Hence  this problem appears to give a 
stimulating testing ground for the Gauge/Gravity correspondence and its physical 
relevance for QCD and particle physics.

Our aim in these lectures is to provide one possible introduction to those 
aspects of the construction of string theory and its applications, mainly the 
AdS/CFT correspondence, which could be of interest for the students in QCD and 
QGP phenomenology. The presentation is thus ``strong-interaction oriented'', 
with both 
reasons that it uses as much as possible the particle language, and that the 
speaker is more appropriately considered as a particle physicist than a string 
theorist. In this respect he is deeply grateful to his string theorists 
friends and collaborators, in first place Romuald Janik, for their 
help in many subtle and often technical aspects of string 
theory. In this respect,
it is quite stimulating to take part in   casting a new bridge between 
``particles and strings''.

\vspace{.2cm}  

{\bf \numero{2}}{\it The Veneziano Formula and Dual\ Resonance\ Models}

\vspace{.2cm} 

\begin{figure} [hbt]
\epsfig{file=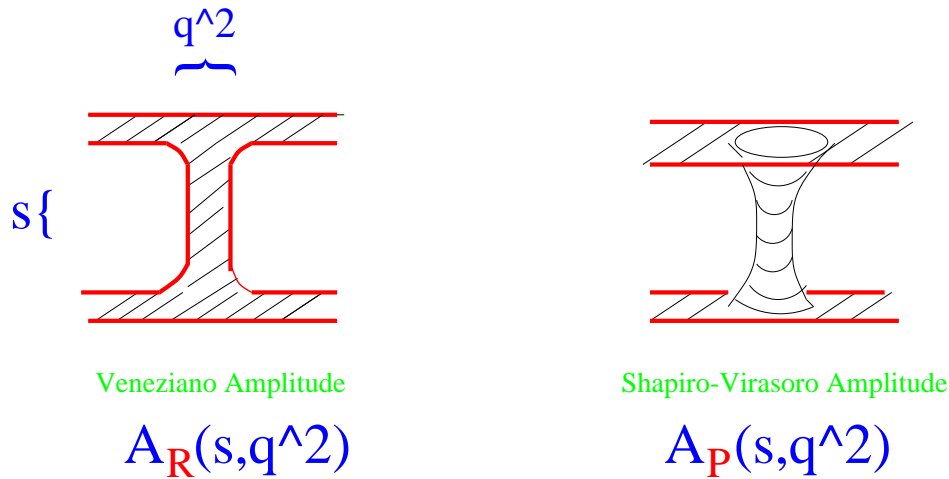,width=12.5cm}
\caption{{\it ``Duality Diagrams'' representing the Veneziano and 
Shapiro-Virasoro 
amplitudes}. The hatched surface gives a representation of  the string 
worldsheet of the process. $s$ (resp. $q^2$) are the square of the c.o.m. energy 
(resp. 
momentum 
transfer) of the 2-body scattering amplitude. $A_{\cal R}(s,q^2)$ is, at high 
energy, the Reggeon amplitude; $A_{\cal P}(s,q^2)$ is the Pomeron amplitude, 
see 
text.}
\label{dual}
\end{figure}
The Veneziano amplitude was the effective starting point of string theory, even 
if 
it took some years to fully realize the connection. At first the Veneziano 
amplitude was proposed as a way to formulate mathematically an amplitude which 
could describe a troubling phenomenological feature of two-body hadron-hadron 
reactions; the ``Resonance-Reggeon'' duality. Indeed, the prominent feature in 
the 
 low-energy domain of two-body hadron-hadron reactions is the presence of 
numerous 
hadronic resonances, such that in some channels one can even describe the whole 
amplitude as a superposition of resonances. At high-energy, the  two-body 
hadron-hadron reactions can also be described by the combination of amplitudes 
corresponding to the exchange in the crossed channel ($q^2$ channel in 
Fig.~\ref{dual}) 
of   particles and resonances, under the 
form of  Regge poles which correspond to an analytic 
continuation in the spin variable. 

The term ``duality'' has been introduced to characterize the fact 
that one should not describe the amplitude by adding the two kinds of 
description. 
On 
the contrary, one expects an equivalent description in terms of a superposition 
of 
resonance states and as a superposition of Reggeon contributions. In order to 
represent this feature, ``duality diagrams'' have been proposed, as shown 
in Fig.~\ref{dual}, where the 2-dimensional surface is drawn to describe the 
summation over states in the direct channel ($s$-channel resonances, whre $s$ is 
the 
square of the total energy) as well in 
the 
exchanged one ($t$-channel reggeons where $t=-q^2$ is the analytic continuation 
of the 
square of the total energy in the crossed channel). In terms of strings, 
it will correspond to the string worldsheet associated to the amplitudes. 

The 
phenomenological 
constructions which were proposed to formulate this property are called the 
``Dual Resonance Models''. As we shall see further, this qualitative 
representation will be promoted to a rigorous  meaning in terms of string 
propagation and interaction. Note already that a topological feature emerges 
from 
the diagrams of Fig.~\ref{dual}. Indeed, if one closes the quark lines (in red 
in Fig.~\ref{dual}) corresponding to the ingoing and outgoing states, they are 
characterized by a planar 
topology 
(Reggeon exchange, left diagram in Fig.~\ref{dual}) or a sphere topology with 
two holes (Pomeron exchange, right  diagram in Fig.~\ref{dual}). This 
topological 
characteristics are indeed a basic feature of string theory, corresponding to 
the 
invariance 
of the string amplitudes w.r.t. the parametrization of the surface spanned by 
the 
string.

\vspace{.3cm} 
{\bf [Exercise 2.1: Show that the ``Shapiro-Virasoro diagram'' is topologically 
equivalent to sticking two ``Veneziano diagrams'' together in a specific way, 
$i.e.$ 
with a ``twist'']}
\vspace{.3cm}

The first and pioneering step in the theoretical approach to dual resonance 
models is the proposal by Veneziano of a mathematical realization of the dual 
amplitude corresponding to the planar topology (Reggeon exchange) the well-known 
``Veneziano Amplitude''. In its simplest version it reads:
\beg 
A_{\cal R}(s,t)=\f{\Gamma(-\al(s))\ 
\Gamma(-\al(t))}
{\Gamma(-\al(s)-\al(t))}
\label{veneziano}
\ee
with  $t\equiv -q^2$ and linear ``Regge trajectories''
\beg
\al(m^2) = \al (0) + \alp m^2\ .
\label{trajectory}
\ee
The Veneziano amplitude has quite remarkable features, thanks to the properties 
of the Gamma function, which give an explicit realization of the duality 
properties. Indeed, in the s-channel as well as in the t-channel,  it 
corresponds 
to an infinite  series of poles (and thus of states), but with a finite number 
of spins for 
each value of positive integer ``level'' $\al(m^2)=n$, since
\eqn
 \ \al(s\ or\ t)\to n \in \nn \ &\Rightarrow&\  
{\f{\Gamma(-\al(s))\Gamma(-\al(t))}{\Gamma(-\al(s)-
\al(t))}} 
\ \approx\n
&\approx& \f{Polynomial^{\{degree \le n\}}\ (t\ or\ s)}{(n-\al(s\ or\ t))}\ .
 \label{poles}
 \eqnx
{\bf [Exercise 2.2: prove formula (\ref{poles}) from Gamma function 
properties]}
\vspace{.3cm}

Concerning the high-energy behaviour, one obtains
\beg 
\ s\to\infty \ \Rightarrow \ A_{\cal R}(s,t)={\f{\Gamma(-\al(s))\ 
\Gamma(-\al(t))}{\Gamma(-\al(s)-\al
(t))}} 
\approx \ s^{\al(t)}\ \Gamma(\!-\!\al(t)) \ , 
 \label{regge}
 \ee
which is the typical dominant ``Regge behaviour'', phenomenologically observed 
in 
hadron-hadron reactions, where the high-energy amplitude in the s-channel 
corresponds to the dominant Regge trajectory (higher spin for a given mass) in 
the 
crossed channels. Subdominant terms correspond to secondary linear regge 
trajectories. 
A similar 
approach was proposed for the sphere topology (the ``Pomeron exchange''), 
resulting in the Shapiro-Virasoro amplitude $A_{\cal P}(s,t).$ 

\vspace{.3cm} 
{\bf [Exercise 2.3: prove formula (\ref{regge})  from  Gamma function 
properties]}
\vspace{.3cm}

For phenomenological purpose, despite its remarkable properties, the Veneziano 
amplitude is not the full answer. Among other problems, all poles are on the 
real 
s or t axis, and thus they correspond to unphysical stable states and not 
resonances. In the following we shall focus on the theoretical meaning of the 
Veneziano amplitude as the seed for string theory. As we shall see, a rigorous 
connection between the Veneziano amplitude and strong interaction physics which 
was its initial motivation, required a more sophisticated framework.
 
\vfill
\eject 
\vspace{.1cm}

\vspace{.1cm}  {\bf \numero{3}}{\it From Dual Amplitudes to Strings}
\vspace{.1cm}
\vspace{.1cm}
\vspace{.1cm}

\begin{figure} [hb]
\epsfig{file=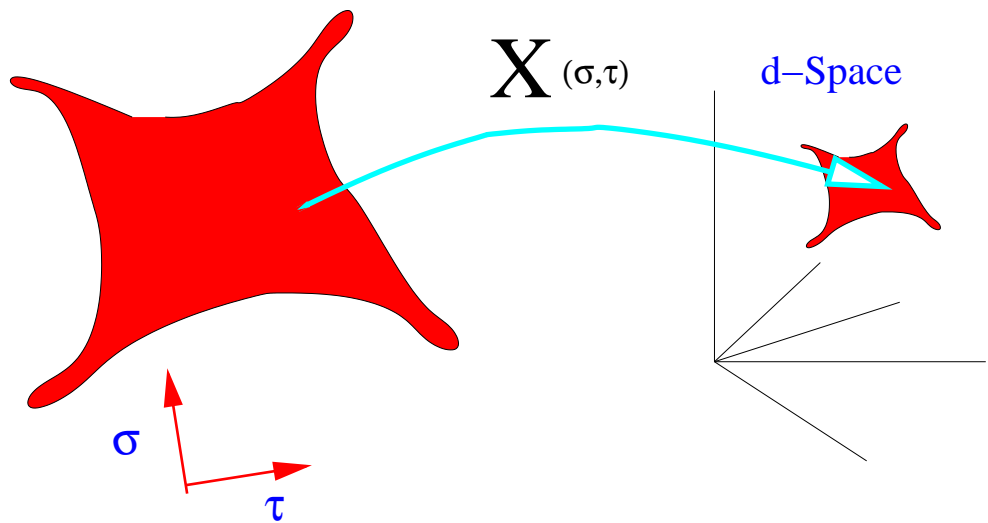,width=13cm}
\caption{{\it String apparatus}. Left: the 2-dimensional $(\sigma,\tau)$ 
string worldsheet. Right: the string embedding in the target space (here: flat 
d-dimensional space). ${\bf X^{\mu}} (\sigma,\tau)$
 is the string position operator, see text.} 
\label{string}
\end{figure}
As clear from formula (\ref{poles}), the Veneziano amplitude  corresponds to an 
infinitely growing number of states as a function of the level 
($n=\alpha(m_n^2)$). Such a spectrum is reminiscent of the classical oscillatory 
modes of a string. However, the construction of a $quantum$ theory of strings 
and 
the identification of the Veneziano amplitude as a particular string 
interaction 
amplitude took some time. In the following we will give the structure of the 
quantum position operator for a (bosonic) string and sketch the derivation of 
the 
Veneziano amplitude as the tree-level string  interaction amplitude.

In order to describe the degrees of freedom of a relativistic string, it is 
useful to introduce the following set of bosonic operators:
 \beg
 [a_{n, \mu} , 
a_{-m, \nu} ] = \eta_{\mu \nu}\ \delta_{nm};\quad
[{\hat{q}}_\mu , {\hat{p}}_\nu ] = i \eta_{\mu \nu}\ ,
\label{oscillator}
\ee
where one considers for the target space, see Fig.~\ref{string}, the 
$d$-dimensional flat metrics
\beg
{\bf \mu,\nu \Rightarrow \eta_{\mu \nu} =
 \left\{1,-1;-1^{*[d-2]}\right\} }\ .
\label{metrics}
\ee
In (\ref{oscillator}), the operators ${\hat{q}}, {\hat{p}}$ describe the 
momentum and position of the string center of mass, while $a, 
a^{\dagger}$ are the bosonic annihilation and creation operators describing the 
oscillator modes of the string. Using these definition, one builds the  string 
position operator at the boundary ${\bf X^{\mu} (\tau,\sigma=0)}$ as follows
\beg
X^{\mu} (\sigma =0,\tau)\equiv {Q_{\mu} (z) = Q^{(+)}_{\mu} (z) 
 +  Q^{(0)}_{\mu}  (z) + Q^{(-)}_{\mu} (z) 
~~;~~z = e^{i \tau}} \ ,
\label{X}
\ee
with
\beg
Q^{(+)} \!=\!  
i \sqrt{2 \alpha'}  \sum_{n=1}^{\infty} \!
\frac{{a_n}}{\sqrt{n}} z^{-n}~;~
Q^{(-)} \!= \! -i \sqrt{2 \alpha'}  \sum_{n=1}^{\infty}\!
\frac{{a_{-n}}}{\sqrt{n}} z^{ n}~;~ Q^{(0)}
\! =\! {{\hat{q}}}\! -\! 2 i  \alpha' {{\hat{p}}} \log z\ .
 \label{Q}
 \ee
The $\sigma$-dependence is restored, specifying the boundary conditions of the 
open string, by multiplying each term $ {a_{\pm n}}/{\sqrt{n}}\ z^{\mp n}$ in  
(\ref{Q})  by $\cos 
n\sigma.$

The calculation of the amplitude is made by integration over the world-sheet 
variables of an overlap over  plane wave operators  
$A \propto \cor{\prod_{j}e^{ip_jX_j}}_{\sigma,\tau}$. Introducing the 
normal ordered vertex operators 
\beg
V(p;z) \equiv :e^{ip\cdot Q(z)}: = e^{ip\cdot Q^{(-)}}\ e^{ip\cdot 
\hat{q}}\ e^{2\alpha' p\cdot {{\hat{p}}}}\ e^{ip\cdot Q^{(+)}} \ ,
\label{vertex}
\ee
the Veneziano amplitude ${B_4}(p_1\!+\!p_2\to p_3\!+\!p_4)$ in terms of 
string vertex 
operators reads:
\beg{
(2 \pi)^d \delta^{(d)}  (\sum_{i=1}^{4}  p_i )\ {B_4} = \int_{0}^{1}
dz_3 
\ \langle 0, p_1 | V(p_2;z_2\!=\!1) V(p_3;z_3)
|
0, p_4 \rangle}\ ,
\label{ampli_4}
\ee
where the external states are 
\beg
\langle 0, p_1 |\propto \langle 0, 0 | V(p_1;z_1\!\to \!\infty)\quad
|0, p_4 \rangle \propto  V(p_4;z_4\!\to\! 0)|0, 0 \rangle
\label{states}
\ee
and where $|0, 0 \rangle$ denotes the vacuum state. The  harmonic oscillators 
acting on 
this state 
build the Hilbert space of string states. The fact that three over the four 
$z_i$ 
coordinates can be fixed at will comes from 
the string symmetries which will be discussed in the next section.
 
 Using the definition (\ref{vertex}) together with the relation
 \beg
 V(p_i;z_i) V(p_j;z_j)=:V(p_i;z_i) V(p_j;z_j): (z_i-z_j)^{2\alp p_i\cdot.p_j}
 \label{relation}
 \ee
 one finally finds
 \beg{B_{4}{=\int_{0}^{1}
  d z}\ z^{-1\!-\!\al(s)} (1\!-\!z)^{-1\!-\!\al(t)} { = 
\f{\Gamma(\!-\!\al(s))\Gamma(\!-\!\al(t))}{\Gamma(\!-\!\al(s)\!-
\!\al(t))}}}
\label{ampli_5}
\ee
which is nothing else han the Veneziano amplitude. An important step towards the 
construction of string theory was made when the suitable generalization to 
arbitrary number of legs $B_4\to 
B_N$ has been performed. The operator formalism was 
then found and fully confirmed.

\vspace{.3cm}
{\bf [Exercise 3.1: prove formula (\ref{relation}) from relations 
(\ref{oscillator},\ref{Q}), using the Baker-Hausdorff formula 
$e^Ae^B=e^Be^Ae^{[A,B]}$ if $[A,B]$ is scalar]}
\vspace{.3cm}

\vspace{.3cm}
{\bf [Exercise 3.2: prove formula (\ref{ampli_5}) from 
(\ref{ampli_4}-\ref{relation})]}
\vspace{.1cm}

\vspace{.1cm} \vspace{.1cm} {\bf \numero{4}}{\it String Symmetries}
\vspace{.1cm} \vspace{.1cm}
 \vspace{.2cm}
 
 The symmetries play a crucial role in the properties of string theory. Let us 
discuss the main features of string symmetries. There exists 
 an exact symmetry group on the string worldsheet. It contains, respectively, 
dilatation, translation and inversion in the worldsheet variable $z\equiv 
e^{i\tau},$ with generators  
$$z \frac{d}{dz}, \quad \frac{d}{dz},  \quad  -z^2  \frac{d}{dz},$$ 
respectively. 
These 
transformations correspond to the infinitesimal generators (the $algebra$) of   
the  Projective 
(conformal) invariance group $SU(1,1)$ (its algebra of generators is given 
further on, see (\ref{su11})). It is precisely this $SU(1,1)$ 
invariance with 3 generators, which allows one to arbitrarily fix $3$ among $N$ 
values of the worldsheet variables in the expression of the amplitude $B_N,$ 
$e.g.$ leaving one interation for the Veneziano amplitude (\ref{ampli_5}).

By extension, one also introduces the  generalized conformal transforms $z^{n+1}  
 \frac{d}{dz},$ for all $n,$ which however will not form an exact symmetry 
algebra at the quantum level, as we will discuss now. They will give rise to the 
famous $Virasoro\ Algebra$ with a ``central extension'' or quantum anomaly.

As usual in the formulation of symmetries, a key point is to find an appropriate 
representation of the algebra in terms of physically meaningful objects, 
here the annihilation and creation operators describing the string. For this 
sake one forms the following operators
\beg
L_n = \sqrt{2 \alpha' n}\ {\hat{p}} \cdot a_{n} \!+\! \sum_{m=1}^{\infty} 
\sqrt{m (n\!+\!m)}\ a_{n\!+\!m} \cdot a_{m}  \!+\!\frac{1}{2} 
\sum_{m\!=\!1}^{n\!-\!1} \sqrt{m ( n\!-\!m)}\ a_{m\!-\!n}\! \cdot\! a_m
\label{Ln}\ee
which possess nice algebraic properties, when acting on the 
string position operator (\ref{X}). 

Let us first consider the set of operators 
$(L_0, L_{-1},L_{1}) .$ One can prove that
\beg
\ {{[ L_0 , Q(z) ] = 
-z \frac{dQ}{dz}~~;~~[L_{-1} , Q(z) ] = -\frac{dQ}{dz}~~;~~
[L_{1} , Q(z) ] = -z^2\  \frac{dQ}{dz}}} \ ,
\label{L}
\ee
which demonstrate that they form an adequate representation of the  algebra of 
the projective symmetry group $SU(1,1).$ More generally one finds 
\beg
[L_{n} , 
Q(z) 
] = -z^{n+1}  
 \frac{dQ}{dz}\ ,
 \label{general}
 \ee
and thus a representation of the generalized projective transformations on the 
string position operator and thus on the string states.

\vspace{.3cm} 
{\bf [Exercise 4.1: prove formulae (\ref{L},\ref{general}) 
using (\ref{Ln}) and the commutation relations (\ref{oscillator})]}
\vspace{.3cm}

Now the symmetry properties will come from the commutation relations between the 
$L_n$ generators, $i.e.$ the Virasoro Algebra. One finds \beg
 \ {[L_n , L_m ] = (n-m) L_{n+m} + 
 \frac{d}{24} n  (n^{2} -1 ) \delta_{n+m,0}}\ .
\label{Virasoro}
\ee

\vspace{.3cm} 
{\bf [Exercise 4.2: prove formula (\ref{Virasoro}), starting with the simpler 
cases when  $n=0,\pm 1.$]}
\vspace{.3cm}

The formula (\ref{Virasoro}) calls for comments. From the algebra, it is easy 
to note that, restricting (\ref{Virasoro}) to $n=0,\pm 1,$ 
one finds
\beg
[L_{\pm1} , L_0] =
 \pm L_{\pm1}~;~[L_{1} ,L_{-1}] = 2L_0
\label{su11}
\ee
which is the algebra of generators of the $SU(1,1)$ group (analoguous 
to $SU(2)$ and its generators $L_\pm,L_z$, but with a change of sign in the 
$[L_{-1} 
, L_0]$ relation which is related to  the non-compactness of the group). Hence 
this 
subalgebra indicates the exact symmetry under projective transforms. 

For higher $|n|> 1,$ one notes the extra contribution $\frac{d}{24} n  (n^{2} -1 
) \delta_{n+m,0},$ which is proportional to the target space dimension $d.$ The 
``central charge'' ($\frac{d}{12}$ with the conventional normalization) 
is a fundamental contribution, showing that the generalized projective group is 
not an exact symmetry (unless other contributions cancel the central 
charge due to the dimension, which is precisely the condition for the existence 
of a 
consistent string theory). It will in fact be crucial in the striking  
feature of string theory to imply a constraint on the target 
space, $i.e.$ the space  in which it moves!
\vspace{.1cm}

\vspace{.1cm}  {\bf \numero{5}}{\it Why 26 (or 10) Dimensions?}

\vspace{.1cm}
\vspace{.1cm}
Now we have the tools to understand the old puzzle which  has jeopardized the 
initial strong-interactions/strings connection. The question is 
whether one can construct  4-dimensional string amplitudes in Minkowski space 
and the answer is in fact ``no''. Let us list the problems when facing the 
construction of 4-d strings in a theoretically consistent way. One finds
th following problems

\vspace{.2cm}
i)
Open (resp. closed) strings $\Rightarrow$ Gauge (resp.   Gravity) at 
lower energy

\vspace{.2cm}
ii) 
{Zero-mass asymptotic states}: gauge bosons, gravitons

\vspace{.2cm}
iii) 
{Hadron spectrum not compatible} 

\vspace{.2cm}
iv) 
{QCD not obtained} 

\vspace{.2cm}
v)
{Problem of dimensions:} The Minkowskian string (resp. superstring) 
target-space is
26 (resp. 10) dimensional

\vspace{.2cm}
{\bf [Exercise 5.1: Given the $\rho$ (spin 1, Mass 770 MeV) and $f_2$ (spin 2, 
mass 1270 MeV) mesons  which are on the dominant hadronic Reggeon trajectory and 
the fact that total hadronic cross-sections are constant with energy (up to 
logarithms) illustrate the third point of the list]}
\vspace{.3cm}
 
Let us consider the problem of dimensions as a major illustration of the   deep 
implications of quantum consistency and symmetries of string theory based on the 
Virasoro Algebra.

The problem can be viewed in different ways. Here we shall take the 
point of view  of the construction of an Hilbert space made of positive 
norm particle states. Let us first remind the well-known 
construction of the Hilbert space for QED. 

If one considers the oscillator construction of the QED Hilbert space one is led 
to satisfy, choosing the covariant gauge, the  condition $$q_\mu 
a^{\dagger}_\mu|0 \rangle  =0\ ,$$ where $q_\mu 
a^{\dagger}_\mu$  denotes the QED analogue (indeed ancestor) of the creation  
L-operator 
$L_1$. 
As is 
known from QED quantification, one may classify the four 
 vector states $a^{\dagger}_\mu|0 \rangle$ within three categories, namely
\eqn
a^{\dagger}_{T}~~|0 \rangle &=& |\phi_{T}
 \rangle\ ~~~\ \!\!\!Transverse~photon~states~~~\sum \vert\phi_{2,3}
 \rangle\vert^2=1 \nonumber \\
a^{\dagger}_{0}-a^{\dagger}_{1}|0 \rangle
 &=& |l \rangle\ \ ~~~~Longitudinal~photon ~state~\ \langle l| l \rangle 
=0\nonumber 
\\
a^{\dagger}_{0}+a^{\dagger}_{1}|0 \rangle &=& |s \rangle
\ ~~~~~Spurious\ photon\ ~state~~~~~~q_\mu a^{\dagger}_\mu \ne 0\ .\nonumber 
\eqnx
In a similar way for strings, and now for the whole hierarchy of operators 
$L_n$, one considers the following (covariant) gauge conditions
\beg
 L_n |\phi_{string}
 \rangle =0\ for\ n>0\ ,
 \label{gauge}
 \ee
which allows a similar generalized classification of states
\eqn
L_n |\phi_{string}
 \rangle&=&0 ~~~~~On\!-\!shell\ states\ ~~~~ positive\ Norm\nonumber \\
\langle l_{string}| l_{string} \rangle &=&0\ ~~~~ Of\!-\!shell\ states\ 
~~~~Zero\!-\!Norm\ ~~~~ 
\nonumber \\
L_n |s_{string}
 \rangle &\ne&0~~~~~Spurious\ states\ ~~~~Unphysical\ States\ .\nonumber 
\label{class}\eqnx
 Now, the key point for the construction of a consistent Hilbert space of string 
states is that spurious states decouple from the other ones. Building a simple 
example we shall prove that it implies a necessary condition over the target 
space dimension. For simplicity, we shall not enter in the proof that this is a 
sufficient condition for eliminating all spurious states from the spectrum. 

Let us consider the following spurious state:
\beg
|s_{string}
 \rangle = L_{-1}|\phi_1
 \rangle + L_{-2}|\phi_2
 \rangle\ .
\label{spurious}
\ee
{\bf [Exercise 5.2: prove that the state defined by formula 
(\ref{spurious})  is indeed spurious if the states $\phi_{1,2}$ are physical 
on-shell states]}
\vspace{.3cm}

Then acting on $|s_{string}\rangle$ with an appropriate combination of 
$L$operators, one finds 
\beg
\left\{L_2 + \f 32 L_{1}L_{1}\right\}
|s_{string}
 \rangle = \sum_i|s_{string},i
 \rangle + \f {{d\!-\!26}}{2} \ |\phi_{string}
 \rangle 
\label{decoupling}
\ee  

{\bf [Exercise 5.3: prove Equation (\ref{decoupling}) by inserting the state 
(\ref{spurious}) using the Virasoro algebra relation  ${[L_2 , L_{-2}] = 4 L_{0} 
+ \frac{d}{4}}$ and classifying the obtained states using (\ref{class})]}
\vspace{.1cm}
 
The decoupling of spurious states requires that the subspace of spurious states 
should remain orthogonal from the physical spaces. Hence one gets the condition 
$d=24,$ characteristic of the bosonic string consistency. A similar condition 
applies to the supersymmetric versions of string theory in Minkowskian space, 
leading to $d=10.$
The decoupling of non-physical states is thus directly a consequence of  the 
Virasoro Algebra and more specifically of its central charge properties.

\vfill
\section*{Lecture II: Gauge/Gravity correspondence}

\vspace{.1cm}  \vspace{.1cm}  {\bf \numero{6}}{\it An Open-Closed String 
connection}
\vspace{.1cm}  \vspace{.1cm}  

We have discussed in section {5} the drawbacks of the initial 
attempts to obtain  strong interaction physics from string theory. 
Indeed, on the 
{\it string-theoretical} point of view, the dimensionality of a Minkowskian 
target space (26 or 10), the existence  of zero mass states and their  
connection to gauge field theory and  
gravity, among other features, seemed to invalidate a string description of 
hadronic interactions. On the {\it 
field-theoretical} point of view, based on the existence of a satisfactory 
theoretical understanding of quark and gluon interactions at weak coupling in 
terms of Quantum Chromodynamics, the challenge of a correct description of  
interactions at long distance relies on the still unsolved problem of computing 
observables at strong coupling. As we shall see now  the Gauge/Gravity duality, 
a deep ``geometrical'' property of string amplitudes, and its precise 
formulation in the case of the so-called AdS/CFT correspondence, seem to 
overcome 
at least in principle, the difficulties from both string and field theory sides 
and opens  a new way for the string approach to strong 
interactions.

\begin{figure}[h]
\includegraphics[width=5.5cm]{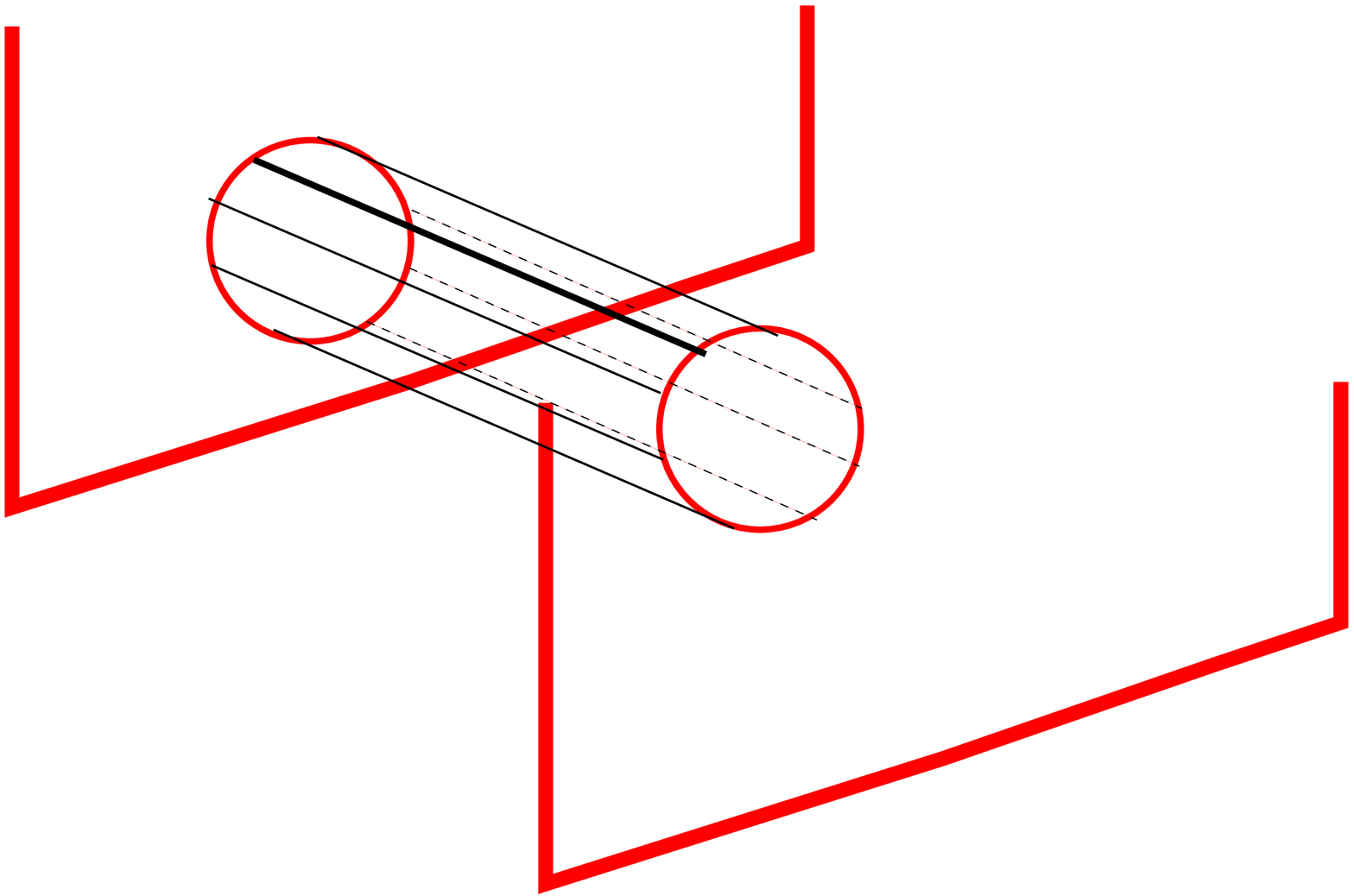}
\hspace{1.5cm}
\epsfig{file=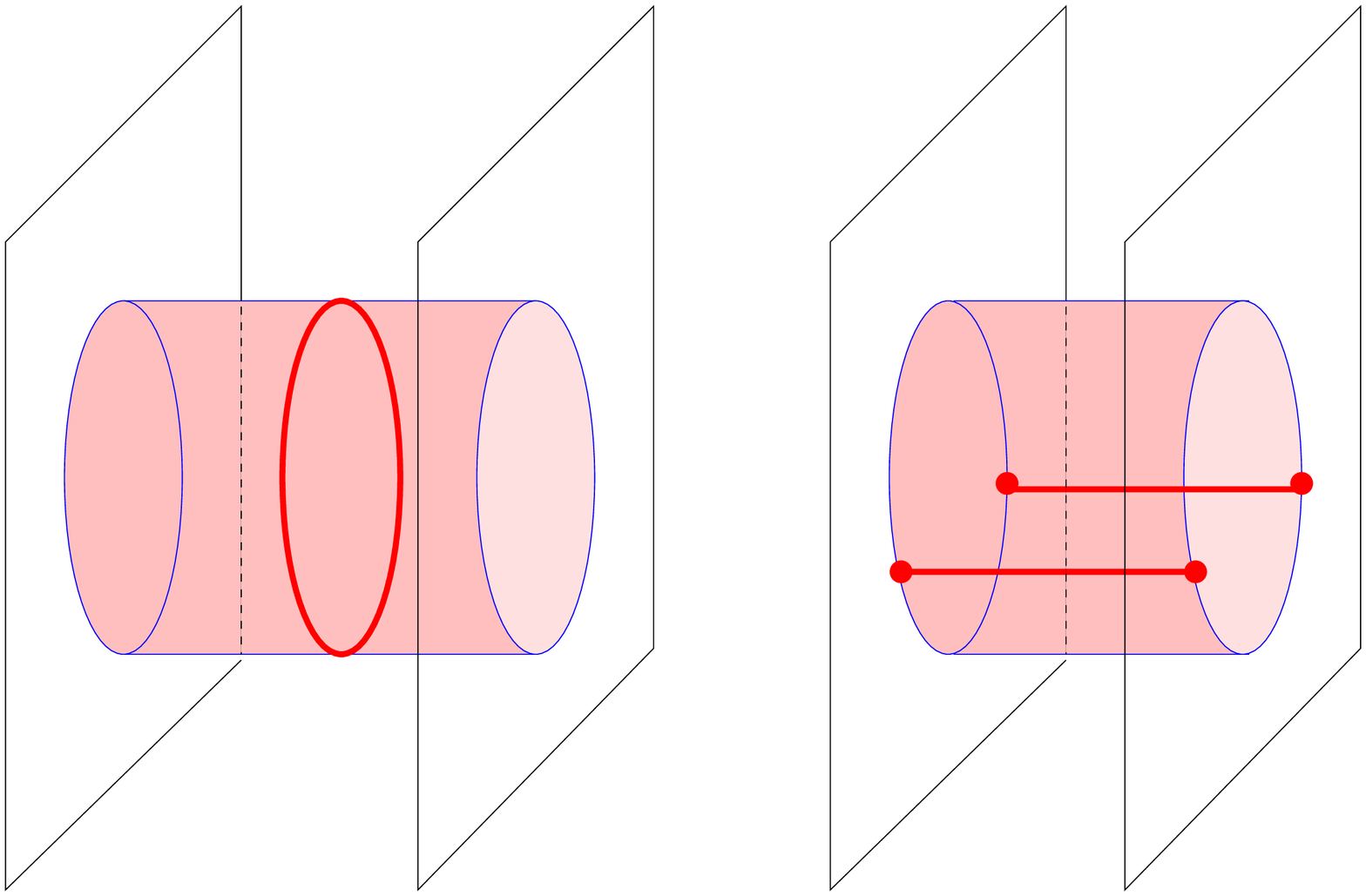,width=5cm}
\caption{{\it  Open $\Leftrightarrow$ Closed duality and  D-Branes.} Left: 
Cylinder topology describing a string  interaction between two stacks of 
D-branes; Right:
the interaction can be described either by the exchange of a closed string 
propagating between the two stacks of branes or by the one-loop contribution of 
an open string attached to the two stacks (from  reference [10]).}
\label{opclo1}
\end{figure}

Let us first give a quite general argument, giving a qualitative explanation of 
this new way of approaching the problem. It relies on a connection between open 
and closed strings which is displayed on Fig.~\ref{opclo1}.
We consider the configuration of two stacks of D-branes in the 
10-dimensional target space of a superstring theory. The D-branes are kind of 
``solitonic'' objects which form a  consistent background in the 
string-theoretical 
framework. In particular, they are the locus of open string endpoints. In 
Fig.~\ref{opclo1} they are understood as stacks of $D_3$ branes which are 
two sets of copies of the $3+1$ Minkowski space, separated by a distance $r$ in 
a fifth dimension, which will play a special role in the following.

One can geometrically interpret the cylinder shape of the interaction in two 
equivalent ways: i) It may be seen as the propagation of a closed loop, starting 
on one $D_3$ brane-stack and reaching the second one; ii) It may be seen 
alternatively as  a one-loop 
contribution from open-strings since open strings may 
have end-points on $D_3$ branes. This equivalence, once given a precise 
formulation in terms of a specific string theory,  has quite intringuing and 
far-reaching consequences. 
  
Let us list some of them:

$\bullet$ Gauge/Gravity duality. As we have alluded to in section {\bf 
{5}}, the massless modes of the string states are gauge fields for 
the open strings and gravitons for the closed strings. Hence the interaction 
amplitude depicted in Fig.~\ref{opclo1} potentially identifies a tree-level 
gravitational interaction with a gauge one at one-loop.

\begin{figure} [ht]
\begin{center}
\epsfig{file=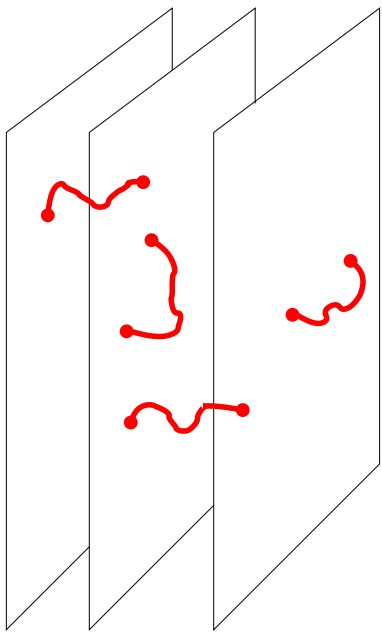,width=5cm}
\end{center}
\caption{{\it $SU(N)$ Gauge theory from $D_3$ branes}. The $D_3$ branes
are considered to be practically at the same location in 10-dimensional space. 
The (short) open string combination of end-points leads to the adjoint 
representation of $SU(N)$.
}
\label{SUN}
\end{figure}

$\bullet$ Short/Long distance relationship. When one consider a large 5th 
dimensional distance $r$, the closed string exchange is 
expected to be described 
by a classical, weakly coupled, gravitational interaction. By contrast, at small 
distance,  the open string interaction is well-described by the exchange of its 
zero-mode states, that is the gauge vector fields. This is theoretically  
justified, since the exchange of open strings with multiple 
combinations of open-string end-points between stacks of near-by $D_3$ branes, 
 are 
related  at weak string coupling to  generic $SU(N)$ gauge field theories (see 
Fig.~\ref{SUN}).

$\bullet$ Weak/Strong coupling relationship. At {\it short distance} $r,$ the 
$SU(N)$ gauge coupling is weak (due to asymptotic freedom for $N\ge 2$). By 
contrast, the gravitational 
interaction is expected to become strong since the stacks of  $D_3$ branes are 
some kind of very massive objets and are expected to generate a strong 
gravitational field in their neighbourhood. On the other end of the comparison, 
at {\it long 
distance} $r,$ the gravity is weak, while the open string 
interaction is expected to become strongly coupled, and moreover, all the 
excitations of the open strings which correspond to the massive oscillator modes  
are expected to contribute. 

From that comes the main feature of the Gauge/Gravity duality; It makes a deep 
connection between weak coupling on one side of the correspondence to the strong 
coupling regime of the other side. It is thus intrinsically a {\it weak/strong 
coupling duality}. 

In the present series of lectures, we are interested in the  weak gravity/strong
gauge coupling combination (the
investigation of the  weak gauge $vs.$ strong gravity
 duality  is yet another fascinating 
challenge). Obtaining valuable new tools of investigation of gauge theories at 
strong coupling from their gravity duals at weak coupling, and thus accessible 
to a quantitative approach.

\vspace{.1cm}

\vspace{.1cm}  {\bf \numero{7}}{\it The AdS/CFT correspondence}

\vspace{.1cm}
\vspace{.1cm}

 \vspace{.2cm}

The AdS/CFT correspondence
  has many interesting both formal and 
physical facets. Concerning 
the 
aspects which are of interest for our problem, it allows one to find 
relations 
between gauge field theories at strong coupling and string gravity at 
weak 
coupling  in the limit of large number of colours ($N_c\!\to\! \infty$). It can 
be examined  quite precisely in  the 
AdS$_5$/CFT$_4$ 
case where CFT$_4$ is the 4-dimensional  conformal field theory  corresponding 
to the $SU(N)$ gauge 
theory 
with ${\cal N} \!=\!4$ supersymmetries.
 
\vspace{.3cm}
{\bf [Exercise 6.1: How many gauge bosons are expected in Fig.~\ref{SUN}?]}
\vspace{.3cm}

Some existing extensions to other gauge theories with  broken conformal 
symmetry
and less or no supersymmetries will be valuable for our approach, since 
they 
lead to confining gauge theories  which 
are 
more similar to  QCD. Note  that the 
appropriate 
string gravity dual of QCD has not yet been identified, and thus we are 
forced 
to restrict for the moment our  use of AdS/CFT correspondence to  generic 
features of  confining theories duals, see a 
discussion 
further on in this section.

\begin{figure}[ht]
{\epsfig{file=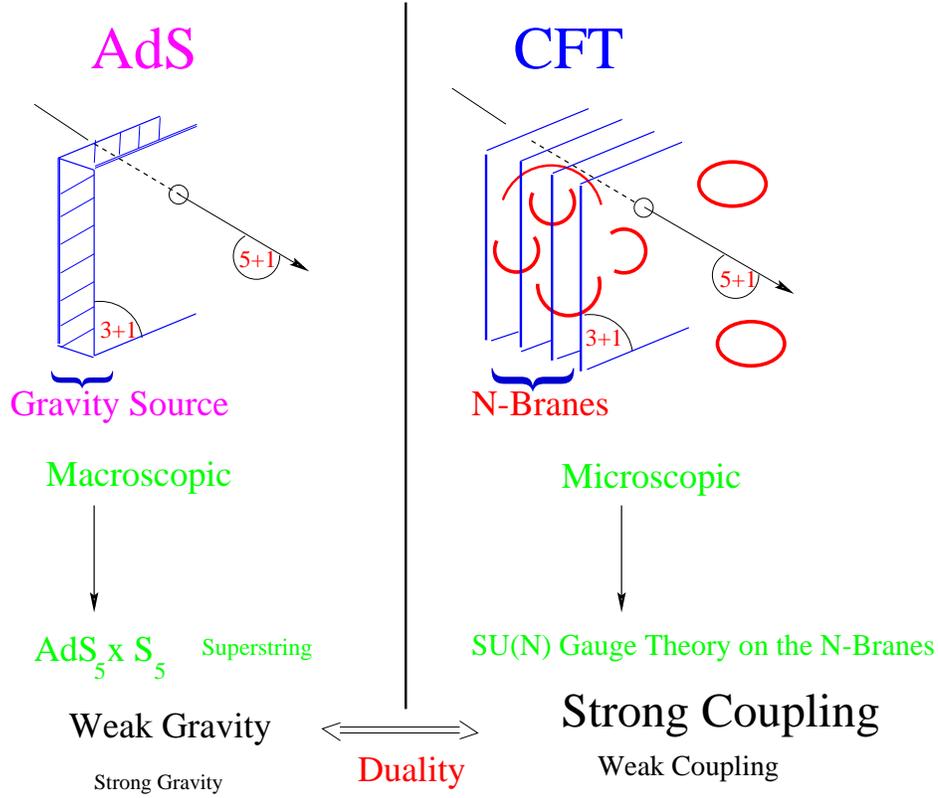,angle=0,width=12.5cm}%
 \vspace{.5cm}
 \caption{{\it Sketch of the AdS$_5$/CFT$_4$ correspondence.} Left: At long 
distance, the gravity source of the branes generate an Anti de Sitter background 
metric; Right: At short distance, the open strings on the branes boil down to a 
non-abelian $SU(N)$ gauge field theory with ${\cal N}=4$ supersymmetries.}%
	\label{2}}
\end{figure}

Let us  recall the canonical derivation leading to  the AdS$_5$ 
background 
see 
Fig.~\ref{2}. One starts from  the (super)gravity classical 
solution of a system of $N\ D_3$-branes in a $10\!-\!D$ space of the (type 
IIB) 
superstrings. 
The metrics solution of the (super)Einstein equations read
\beg
ds^2=f^{\!-\!1/2} (\!-\!dt^2\!+\!\sum_{1\!-\!3}dx_i^2) 
\!+\!f^{1/2}(dr^2\!+\!r^2d\Omega_5) 
\ ,
\label{super}
\ee
where the first four coordinates are on the brane and $r$ corresponds to 
the  
coordinate along the normal to the branes. In formula (\ref{super}), one writes
\beg
f=1+\frac {R^4}{r^4}\ ;\ \ \ \ \ R=4\pi g^2_{YM}\alpha'^2 {N} \ ,
\label{R}
\ee
where $g^2_{YM}{N}$ is the so-called `t Hooft-Yang-Mills coupling equal to the 
string coupling $g_s$ and $\alpha'$ the 
string 
tension.

One considers the   limiting behaviour considered by 
Maldacena, 
where 
one zooms on the  neighbourhood of the branes  while in the same time going to 
the limit of 
weak 
string 
slope $\alpha'.$ The near-by space-time is thus distorted due to the 
(super) 
gravitational field of the branes. One goes to the limit where 
\beg
\ R\ fixed\ ; \frac {\alpha'(\to 0)} {r (\to 0)}\to z\ fixed\ .
\ee
This, from the second equation of (\ref{R}) obviously implies 
\beg
{\alpha'\to 0}\ , \ g^2_{YM} 
{N}\sim \f{R}{4\pi\alpha'^2}\to \infty \ ,
\ee
{\it i.e.} both a weak coupling  limit for the string theory and   a strong 
coupling limit for the dual gauge field theory.
By reorganizing the two parts of the metrics one obtains
\beg
\ {ds^2={ \frac 1{z^2} (-dt^2+\sum^3_{1}dx_i^2+ dz^2)} +  {R^2 
d\Omega_5}}\ ,
\label{AdS}
\ee
which corresponds to the 
{AdS$_5$} $\times \  {S_5}$ background structure. In (\ref{AdS}) $\frac 1{z^2} 
(-dt^2+\sum_{1-3}dx_i^2+ dz^2),$   describes a  Anti de Sitter\footnote{the 5-d 
version of de Sitter geometrical space, whose 4-d version appears in general 
relativity,  has a plus sign, $i.e.$ obeys the equation $-x_0^2 +x_1^2+ 
\sum_{i=2}^6 x_i^2 = R^2 .$.} 
geometrical space which is a 5-dimensional  hyperbolo\"\i d  of equation 
$-x_0^2 -x_1^2+ \sum_{i=2}^6 x_i^2 = -R^2 $ in a 6-dimensional flat Minkowski 
space   
${S_5}$ is  
the 5-sphere of metric ${R^2 
d\Omega_5}$ where $\Omega_5$ is the 5-dimensional solid angle. More detailed 
analysis shows that the isometry group of 
the  
5-sphere may be considered as  the ``gravity dual'' of the ${\cal N}\!=\!4$ 
supersymmetries (for completion, the quantum number dual to $N_c,$ the number 
of 
colours, is the invariant charge carried by a Ramond-Ramond form 
field in (type 
IIB) 
superstrings.

In the case of confining backgrounds, an intrinsic scale breaks conformal 
invariance and is brought in the dual theory  through {\it e.g.} a 
geometrical constraint. For instance,  a 
confining gauge theory is expected to be dual to string theory in an $AdS_{BH}$ 
Black
Hole (BH) background 
\beg
\label{e.bhmetric}
ds^2_{BH}=\f{16}{9}\f{1}{f(z)}\f{dz^2}{z^2} + \f{\eta_{\mu\nu}dx^\mu
dx^\nu}{z^2} + \ldots 
\ee
where  $f(z)=z^{2/3}[1-(z/R_0)^4]$ and $R_0$ is the position of 
the BH horizon.
One may  use this type of 
background 
to
study the interplay between the confining nature of gauge theory and
its reggeization properties. Actually the qualitative 
arguments and
approximations should be generic for most confining backgrounds.
For instance, 
other geometries for (supersymmetric)
confining theories
 have been discussed in this respect. 
They have the property that for
small $z$, the geometry looks like 
$AdS_5\times
S^5$ (in 
 up to logarithmic corrections related to asymptotic
freedom) giving a coulombic $q\qb$ potential. For large $z$ the
geometry is effectively flat. In all cases there is a scale, similar
to $R_0$ above, which marks a transition between the small $z$ and
large $z$ regimes.
\vspace{.1cm} 

\vspace{.1cm}  {\bf \numero{8}}{\it Wilson Loops,  Minimal  Surfaces and 
Confinement}

\vspace{.1cm}  
\vspace{.1cm}

We will find 
appropriate in the next section to formulate the scattering amplitudes in terms 
of 
Wilson loops, 
since the Gauge/Gravity ``dictionnary'' for Wilson loops  has been 
proved to be well suitable for dual properties in general. Let us thus  
introduce this 
dictionnary in the following. 
Let us introduce the general framework within which Wilson loops in  the 
``boundary'' 
gauge field theory are in correspondence with minimal surfaces in the ``bulk''.

\begin{figure}[hbt]
\epsfig{file=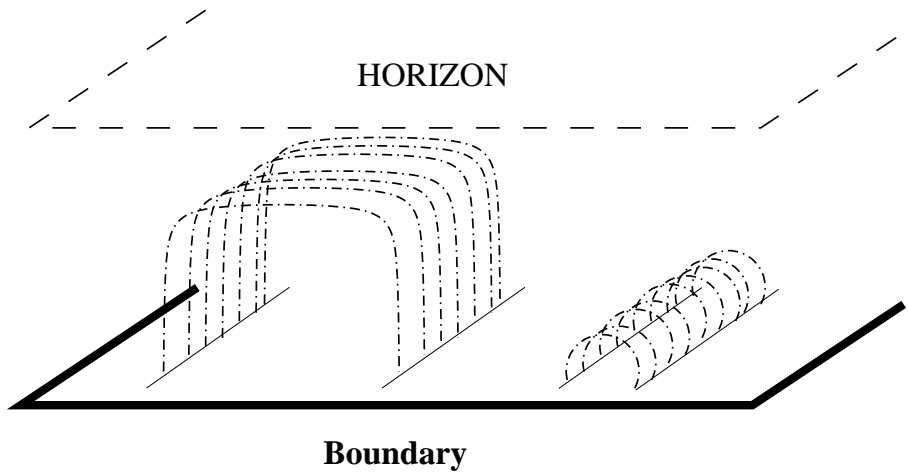,angle=0,width=12cm}%
 \caption{{\it Sketch of minimal surfaces with Wilson loop boundaries: the 
potential 
configuration.} The Wilson loops correspond here to the potential configuration 
with 
transverse boundaries at distance $L$ and parallel time-like boundaries going to 
infinity at the limit, 
see text. Left:  Minimal surface in the presence of a confining background 
such as (\ref{e.bhmetric}); Right: Minimal surface corresponding to an 
$AdS_5$-like  
background such as (\ref{AdS}). 
NB: In the figure, it is represented in  an approximate  case when the ratio of 
the 
Wilson loop transverse size to the horizon is small. Without horizon, the 
minimal 
surface at large transverse size would extend without limit.}\label{f3}
\end{figure}

In a  framework suitable for performing the AdS/CFT correspondence, quarks 
(resp. antiquarks) can be represented\footnote{In order to get correct quark 
degrees 
of freedom, $e.g.$ flavor, a more complete geometrical set-up including $D_7$ 
branes 
is used.} by  colour sources in the fundamental  
(resp. 
anti fundamental)
reps. of SU(N). In order to illustrate the way how one may formulate in practice 
the 
AdS/CFT 
correspondence in a context similar to QCD, 
let us consider first the example of  the vacuum expectation value ({\it vev}) 
of 
Wilson loops in a  configuration parallel
 to the time direction of the branes. We consider the large time limit and thus 
the 
loops close ``near'' infinity in the time direction (see Fig.~\ref{f3}). This 
configuration
allows 
a   determination of the 
potential between colour charges. The r\^ole of colour charges
 is played by open string states 
elongated 
between a  stack of $N_c$ $D_3$ branes on one side and one $D_3$ brane 
near the 
boundary of AdS space.

The correspondence can be formulated\footnote{For simplicity, an extra singlet 
term 
in the left-hand exponent, allowing to cancel divergences, is not written here.} 
as follows
\beg
\ {\langle\ e^{iP \int _C \vec A\cdot\vec dl}\ \rangle \ 
=\ \int^{^{}}_{\Sigma}\ e^{- 
\frac{Area({\Sigma})}{\alpha'}}}\ ,
\label{Wilson}\ee
where $C$ is the Wilson loop contour near the $D_3$ branes  and $\Sigma$ 
any surface in $AdS$-space with $C$ as the boundary, see Fig.~\ref{f3}. 

In the semi-classical approximation for the right-hand (gravity) side of the 
relation 
(\ref{Wilson}) where the Gauge/Gravity  correspondence would give the strong 
coupling 
value of the left-hand (gauge) side,  the integration over surfaces $\Sigma$ 
boils down to
\beg
\langle\ e^{iP\int_C\vec A\cdot\vec dl}\ \rangle \approx e^{-
\frac  {Area_{min}}{\alpha'}}\times 
Fluct.\ ,
\label{Minimal}\ee
where  $Area_{min}$ is the minimal surface whose boundary is the gauge-theory 
Wilson 
loop.   The factor denoted $Fluct.$ refers to the  fluctuation 
determinant  
around the minimal surface, corresponding to the first  quantum correction 
beyond the 
classical approximation. It gives  an interesting  calculable semi-classical 
correction, as we shall see on the amplitude exemple.

  In  Fig.~\ref{f3}, we have 
sketched  the form of  minimal surface solutions for the  
``confining'' $AdS_{BH}$
case,   (see 
above (\ref{e.bhmetric})). For large separation of Wilson lines, the minimal 
surface is bounded near the horizon and is consequently curved. At smaller 
separation, 
the solution becomes again similar to the conformal case, since the horizon 
cut-off does not play a big r\^ole.

In gauge theory, the quark-quark potential is known to be obtained from a 
suitable 
time-like infinite limit of the rectangular Wilson loop {\it vev}.  One has 
\beg
\ {V(L)= \lim_{T \to \infty} \f 1T \times  \log \langle\ Wilson\ Loop\ \rangle
}\label{potential}
\ee
Thanks to the Gauge/Gravity correspondence (\ref{Wilson}) and the  classical 
approximation (\ref{Minimal}), one is able to get the strong coupling limit of 
the 
interquark potential  from the large time limit of the Wilson loop v.e.v.:
\begin{eqnarray}
AdS_5: \langle Wilson\ Loop\rangle&=&e^{TV(L)}\sim 
e^{\#_1\sqrt{g^2_{YM}N}\ \f TL}\nonumber \\ AdS_{BH}: \langle Wilson\ 
Loop\rangle&=&e^{TV(L)}\sim e^{\#_2\ \f {TL}{R_0^2}}\ , \nonumber 
\end{eqnarray} 
where $\#_{1,2}$ are constants. The potential behaviour obeys the  nonconfining
Coulomb law $V(L)\propto 1/L$ for the $AdS$ case and the confining law 
$V(L)\propto L$ for the $AdS_{BH}$ case. An
interesting  nonanalytic dependence over the square-root coupling  appears. Note 
again that, even in the case of a 
confining 
geometry with an horizon at $R_0,$ Wilson lines separated by a distance 
$L<<R_0 
$ do not give rise to minimal surfaces  sensitive to the horizon (see 
Fig.~3), 
and thus  to classical solutions similar to the non-confining case, 
which can 
give interesting indications for small spatial separation.

The important r\^ole of  fluctuation corrections and the way of computing 
it 
in some non-trivial cases will be discussed further on.

\vspace{.1cm}

\vspace{.1cm}  {\bf \numero{9}}{\it Application:  A dual model for  Dipole 
Amplitudes}
\vspace{.1cm}
\vspace{.1cm}

\begin{figure}[ht]
\epsfig{file=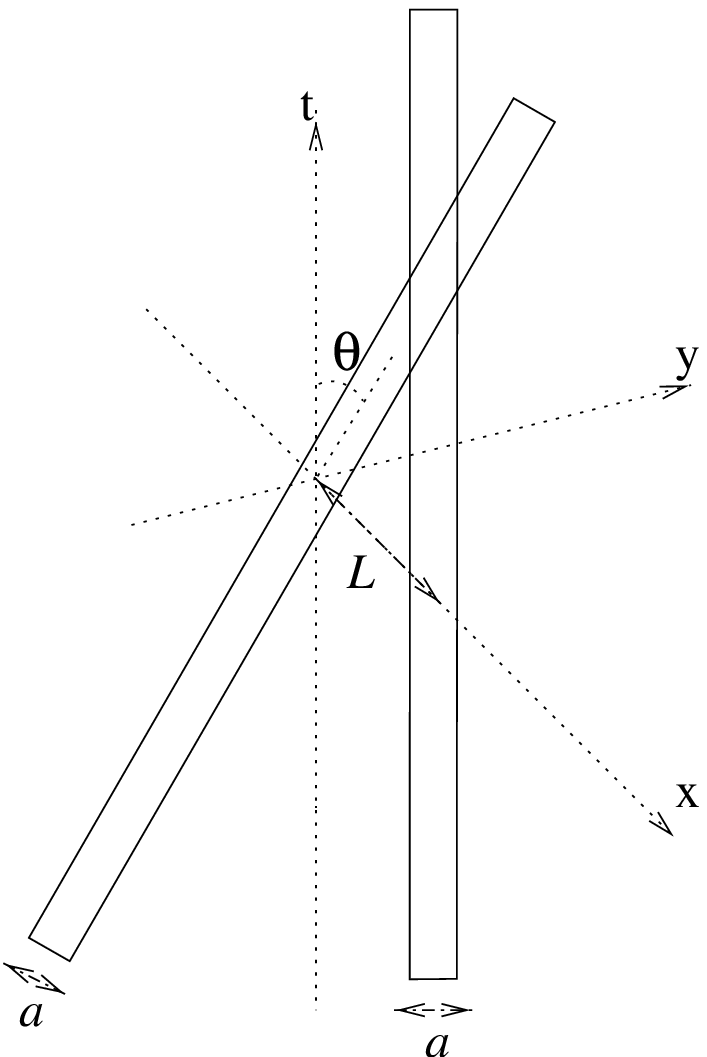,angle=0,width=3.5cm}
\hspace{2cm}
\epsfig{file=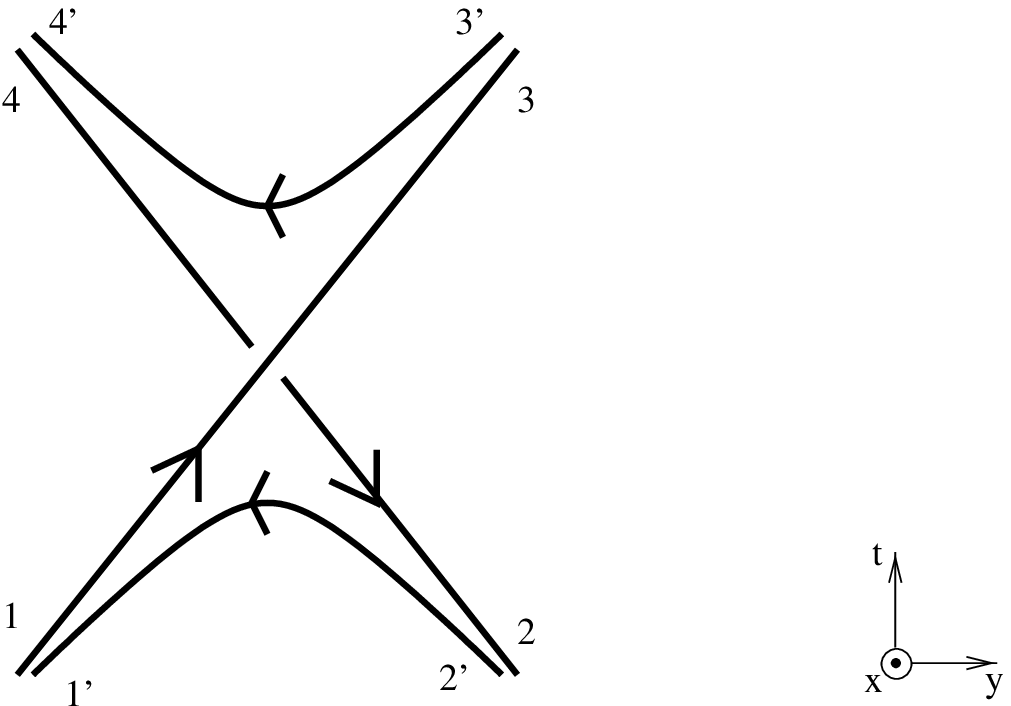,angle=0,width=7cm}
    \caption{{\it Wilson loops for  Dipole-Dipole scattering in configuration 
space}. 
The figure is 
drawn in the physical  {\it boundary} configuration space $(t,x,y,z).$
 Left: the two Wilson loops corresponding to the elastic dipole-dipole 
amplitude 
$A_{\cal P}^{d\!-\!d}(s,q^2) . $ $L$ is the impact parameter distance between 
the 
colliding dipoles, $a$ is the (small) $q\bar q$ distance  in the dipoles. All 
$q\bar q$ trajectories are straight lines in the  eikonal  approximation
Right: the Wilson loop $(1\! \to\! 3 \! \to \!3' \! \to \!
4' \! \to \! 4 \! \to \! 2 \! \to \! 2' \! \to \! 1' \! \to \!1)$  in 
configuration space corresponds to the 
inelastic dipole-dipole amplitude  $A_{\cal R}^{d\!-\!d}(s,q^2).$  The Wilson 
lines 
$1\! \to \! 3$ and $4 \!\to\! 2$ are taken as straight lines in the eikonal  
approximation 
(see text).}
 \label{fig}
\end{figure}
Now we will come back to our original problem of describing in a consistent way 
the strong interaction two-body amplitudes which correspond, $e.g.$ to the 
processes depicted in fig.\ref{dual}. There are different  approaches to 
scattering 
amplitudes using gravity duals, including recently the formulation of gluon 
amplitudes 
at strong coupling in the $SU(N)$ gauge 
theory 
with ${\cal N} \!=\!4$ supersymmetries. However, since we are interested in the 
present lecture in the approach to hadronic  scattering amplitudes, we
search for  both a 
field-theoretical formulation based on QCD and the determination of the gravity 
duals of the corresponding amplitudes.  Concerning the nature of the 
dual theory, the gravity dual theory of QCD has  not  yet been 
identified. More generally, the problem of deriving a correspondence for a 
confining 
theory with asymptotic freedom is not yet achieved. In the following we shall 
use an 
approach where 
only generic features of confining backgrounds allow to determine some 
properties of the amplitudes. The price we pay is that we will only be  able to 
discuss 
the high-energy behaviour of the amplitudes, $i.e.$  the high-energy regime, 
which was 
discussed in section  2 for instance in Eq.(\ref{regge}). Other properties 
of the amplitudes will not be discussed, and probably are more difficult to 
derive in the absence of a better determined dual background to QCD. 
Using the AdS/CFT correspondence, we will find that two-body high-energy 
amplitudes 
in gauge field 
theories can 
be 
related to specific configurations of 
minimal surfaces.

Using the   Wilson loop properties, 
it is now possible to formulate the Gauge/Gravity correspondence for the  
elastic and 
inelastic scattering 
amplitude of massive colorless $q \bar q$ states in the space of    QCD color 
dipoles. In Fig.~\ref{fig}, one displayed the elastic and inelastic amplitudes 
of two dipoles in configuration space, corresponding respectively to $A_{\cal 
P}^{d\!-\!d}(s,q^2)$, and  $A_{\cal R}^{d\!-\!d}(s,q^2)$ appearing in Lecture I. 
We will 
here 
consider the
amplitudes at high energy, i.e. the problem of ``Reggeization''. Indeed, at high 
energy, fast moving colour sources 
propagate 
along
linear trajectories in coordinate space thanks to the eikonal 
approximation. This important property of high energy propagation of color 
sources will be  helpful for the evaluation of the amplitudes through 
Gauge/Gravity duality. 

Let us first consider the elastic dipole amplitude, $i.e.$  the left diagram of 
Fig.~\ref{fig}. In the gauge field theory, one may write it in terms of a 
correlator 
between two Wilson lines in configuration space, namely
\beg 
\label{e.ampinit}
\ {A_{\cal P}^{d\!-\!d}(s,q^2)=-2is \int d^2\xpr \ e^{iq\xpr}
\cor{\f{W_1W_2}{\cor{W_1}\cor{W_2}}-1}}
\ee
where the Wilson loops ${W_1},{W_2}$ corresponding to the two colliding dipoles 
follow classical straight lines for 
quark(antiquark)
trajectories  and close at infinite time, as for the potential. The 
normalization $\{\cor{W_1}\cor{W_2}\}$ of the correlator ensures that the 
amplitude vanishes when 
the Wilson loops get decorrelated at large distances.

One useful technique is to formulate the duality property in Euclidean  ${\cal 
R}^4$ 
space where it takes the form of  a well-defined geometrical interpretation in 
terms 
of a minimal 
surface 
problem. Then the analytic 
continuation 
from Euclidean to  Minkowski space allows one to find the physical solution.

The Wilson line {\it vev} can be expressed as a minimal surface problem with 
(approximately) 
two copies (for dipole size $a\sim 0$) of a minimal surface
whose 
boundaries are  straight lines in a 3-dimensional coordinate space, 
placed at 
an impact parameter distance $L$ and rotated one with respect to the  other by  
an 
angle 
$\theta,$ 
see 
Fig.~\ref{fig}. then the amplitude will be obtained through the analytic 
continuation
\beg
\theta \leftrightarrow -i\chi \quad ;\ \quad t_{Eucl} \leftrightarrow \ 
it_{Mink}\ ,
\ee
where $\chi = \log s/m^2$ is the total rapidity interval. 

  In flat space, with the same boundary conditions,  the 
minimal surface is 
the 
{\it helicoid}. One thus  realizes that the problem can be formulated as a 
minimal surface problem whose  mathematically well-defined   solution   is 
a 
{\it generalized  helicoidal} manifold embedded in curved background 
spaces, 
such as Euclidean AdS Spaces. Unfortunately, this problem is rather 
difficult 
to 
solve analytically, even in flat space.  It is known as the Plateau 
problem, 
namely the  determination of minimal surfaces for given boundary conditions.

In fact,  the 
definition of the  minimal surface geometry in the conditions of a 
confined 
$AdS_{BH}$ metrics (\ref{e.bhmetric}) appears to be  simpler, at least for the 
leading 
contribution. Indeed, in the 
configuration 
of Wilson lines of Fig.~\ref{f3} in the context  of a confining theory, the  
AdS$_{BH}$ metrics is characterized by a singularity at $z=0$ which 
implies a 
rapid growth in the  $z$ direction 
towards the D$_3$ branes, then stopped near the horizon at $z_0.$ Thus, 
to a good approximation, and for a large enough
impact 
parameter (compared to the horizon distance), the main contribution to 
the 
minimal area is from the  metrics in the bulk near $z_0$ which is nearly
flat. Hence, near $z_0,$ the relevant minimal area can be  drawn on a
{\it classical helicoid}, whose analytic expression is known. This expression 
contains a logarithmic singularity in terms of kinematical variables, which 
turns out to be essential to generate an imaginary part in the action after 
analytic continuation to Minkowski space.

After analytic continuation,  one obtains
\beg 
\label{e.ai}
A_{\cal P}(s,q^2)= 2is\int d\vec l\ e^{i\vec q\cdot \vec l-\left\{ 
\f{\sqrt{2g_{YM}^2N}}{\chi} \f{L^2}{2R_0^2}\right\}}  \propto\ 
s^{1-\f{q^2\ R_0^2}{\sqrt{8g_{YM}^2N}}}\ .
\ee
which represents a  Reggeized elastic 
amplitude, with a linear Regge trajectory 
\beg 
\label{Ptrajectory}
\alpha_{_{\cal P}}(q^2)= \alpha_{_{\cal P}}(0)-q^2\alef_{_{\cal P}}  \equiv 
1-\f{q^2\ R_0^2}{{\tiny{\sqrt{8g_{YM}^2N}}}}
\ee
characterized by a Pomeron
intercept $\alpha_{_{\cal P}}(0)=1$ 
and a Regge slope, defined in terms of the gravity dual  parameters by 
$\alef_{_{\cal P}}=\f {R_0^2}{\sqrt{8g_{YM}^2N}},$ where $g_{YM}^2N\equiv g_s$ 
is 
the 
string or `t 
Hooft coupling. 

Let us now consider the dipole-dipole inelastic amplitude
\beg 
\label{e.scat}
(11')+(22') \lra (33')+(44')\ ,
\ee 
represented in configuration space on Fig.~\ref{fig}, right.
The helicoidal geometry remains  
valid 
due 
to the eikonal approximation for the ``spectator quarks'', namely the $1\!\to\! 
3$ 
and $4 \!\to\! 2$ Wilson lines
while the ``exchanged quarks'' define a trajectory drawn on the 
helicoid. This trajectory plays the r\^ole of a dynamical time-like  
cut-off which takes part in the 
minimization 
procedure. The 
resulting amplitude reads:
\beg 
A_{\cal R}(s,q^2)=2i\int d\vec l\ e^{i\vec q\cdot \vec l-\left\{ 
\f{2\sqrt{2g_{YM}^2N}}{\chi} \f{L^2}{R_0^2}\right\}}\propto
s^{-\f{2q^2\ R_0^2}{\sqrt{2g_{YM}^2N}}}\ ,
\label{reggeon}
\ee
corresponding to a linear Regge trajectory
\beg 
\label{Rtrajectory}
\alpha_{_{\cal R}}(q^2)= \alpha_{_{\cal R}}(0)-\alef_{_{\cal R}}\ q^2 \equiv 
-\f{q^2\ 2R_0^2}{\sqrt{2g_{YM}^2N}}
\ee
characterized by a ``Reggeon'' 
intercept $\al_{_{\cal R}}=0$ 
and a Regge slope $\alef_{_{\cal R}}=\f{2R_0^2}{\sqrt{2g_{YM}^2N}}.$ Note that 
the 
slope $\alef_{_{\cal R}}$ is related to the quark potential calculated within 
the same 
AdS/CFT 
framework and, quite interestingly we find $\alef_{_{\cal R}}=4\alef_{_{\cal 
P}}$, to be compared with the  string result  at weak  coupling $\alef_{_{\cal 
R}}=2\alef_{_{\cal 
P}}.$


Up to now, we restricted ourselves to  a classical approximation based on the  
evaluation of minimal surfaces solutions for the various Wilson loops involved 
in the preceeding calculations. It is interesting to note 
 that a 
further step can be done by evaluating the contribution of quadratic 
fluctuations of the
string worldsheet around the minimal surfaces in the case where these surfaces 
are embedded in helicoids, as discussed for the confining backgrounds.
The semi classical correction   comes from the fluctuations near the 
minimal surface. The main outcome is that this semi classical correction can be 
computed and   is intimately related to the well-known ``universal'' 
L\"uscher 
term contribution to the  interquark potential.
 
After some non-trivial steps, the formulae (\ref{e.ai},\ref{reggeon}) get 
corrected as follows
\beg 
A_{\cal P}(s,q^2) \propto s^{\al_{\cal 
P}(-q^2)}=s^{1+\f{n_\perp}{96}-q^2\f{\alef_{_{\cal R}}}{4}}
\quad   
A_{\cal R}(s,q^2) \propto s^{\al_{\cal R}(-q^2)}\ =s^{\f{n_\perp}{24}-q^2 
\alef_{_{\cal R}}}
\ee
where $n_\perp$ is the number of zero modes of the gravity dual theory in the 
transverse-to-the-branes 
directions.
The result is just equivalent to the known  L\"uscher term in $4d$ found in the 
large-time expansion of 
the rectangular Wilson loop, except that the number of zero modes $n_\perp = 
d-2$ can 
be larger than 
the  usual value (=2) corresponding to  flat $4d$ space.

It is interesting to note that
this theoretical feature is in qualitative agreement with the phenomenology of
soft scattering. Indeed once we fix the $\alef$ from the
phenomenological value of the static $q\qb$ potential ($\alef\sim 0.9\,
GeV^{-2}$) 
we get for the slopes $\al_R=\alef\sim 0.9\, GeV^{-2}$ and
$\al_P=\alef/4\sim 0.23\, GeV^{-2}$ in good agreement with the phenomenological
slopes.

A second feature is the relation between the Pomeron and Reggeon
intercepts. At the classical level of our approach these are
respectively $1$ and $0.$ Note that this classical contribution matches what is 
obtained  from simple exchanges of two
gluons and quark-antiquark pair, respectively, in the $t\equiv-q^2$ channel. The 
fluctuation (quantum) contributions to the Reggeon and Pomeron are also
related by the factor $4$. 
  
Adding both classical and fluctuation contributions gives an estimate
which is in qualitative agreement with the observed intercepts.  
Indeed, when calculating the fluctuations around 
a minimal surface near the horizon in the
BH backgrounds there could be  $n_\perp=7,8$ massless bosonic modes.  For 
$n_\perp=7,8$ one gets $1.073-1.083$ for the Pomeron and 
$0.3-0.33$
for the Reggeon. 
This result is in agreement with the observed intercept for the ``Pomeron'' and 
somewhat below the intercepts  $\sim 0.5$ observed for the
dominant Reggeon trajectories. 
The interesting output of the application 
of 
AdS/CFT correspondence to high energy amplitudes at strong coupling is 
to 
emphasize the relation between Reggeization and confinement, using the 
description of two-body scattering amplitudes in the dual string theory.
\vfill
\eject
\section*{Lecture III: Quark-Gluon Plasma/Black Hole Duality}

\vspace{.1cm}
\vspace{.1cm}  {\bf \numero{10}}{\it QGP formation and Hydrodynamics}
\vspace{.1cm}\vspace{.1cm}
\vspace{.5cm}
\begin{figure}[hb]
\begin{center}
\includegraphics[width=30pc]{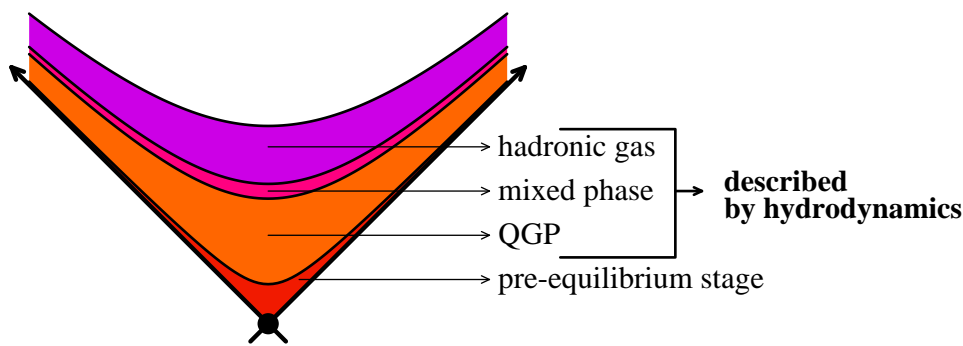}
\end{center}
\caption{{\it Description of QGP formation in heavy ion collisions}. The 
kinematic landscape is defined by ${\tau = \sqrt{x_0^2-x_1^2}\ ;\ {\eta=\f 12 
\log \f 
{x_0+x_1}{x_0-x_1}}\ ;\ {x_T\!=\!\{x_2,x_3\}}}\ ,$
where the coordinates along the light-cone are $x_0 \pm x_1,$ the transverse 
ones 
are $\{x_2,x_3\}$ and $\tau$ is the proper time, $\eta$ the ``space-time 
rapidity''.}
\label{QGP}
\end{figure}

The formation of a QGP (Quark Gluon Plasma) is expected to be realized in 
high-energy 
heavy-ion collisons, {\it e.g.} at RHIC and soon at the LHC. One of the main 
tools
for 
the description of such a formation is the relevance of relativistic 
hydrodynamic 
equations in some intermediate stage of the collisions, see Fig.~\ref{QGP}.
The problem of the hydrodynamic description is the somewhat indirect   relation 
with the 
underlying 
fundamental theory. Indeed, the experimental observations seem to indicate an 
almost 
perfect-fluid behaviour with small shear viscosity, which naturally leads to 
consider a 
theory at strong coupling and thus within the yet unknown non-perturbative 
regime 
of 
QCD. Moreover the QGP formation appears to be fast, which may
also 
point towards strong coupling properties. Another key point of the standard 
description is the approximate boost-invariance of the process in the central 
rapidity 
region, that is the well-known {\it Bjorken flow}. The goal of the string 
theoretic 
approach is 
to make use 
of the Gauge/Gravity correspondence as 
a 
way to tackle the problem of the hydrodynamic behaviour from the fundamental 
theory 
point of view. It allows to draw quantitative  relations
between a strongly coupled gauge field theory and a weakly coupled string theory

More specifically, the AdS/CFT correspondence between the ${\cal N}\!=\!4$ 
supersymmetric 
$SU(N)$ gauge theory and  superstrings in 10 dimensions can be used as a 
calculational 
laboratory for this kind of approach, at least as a first stage before a more 
realistic 
application to QCD. The unconfined character of the QGP gives some hope that the 
explicit AdS/CFT  example could be useful despite
the lack of asymptotic freedom and other aspects specific of QCD.

\vspace{.1cm}
\vspace{.1cm}  {\bf \numero{11}}{\it AdS/CFT and Holographic  Hydrodynamics}
\vspace{.1cm}\vspace{.1cm}

One typical and fascinating aspect of the Gauge/Gravity duality is the property 
of 
$holography$ as we have seen in section  8. It states that the amount of 
information contained in the boundary 
gauge 
theory (on the brane) is the same as the one contained in the bulk string 
theory. 
In our 
problem, we shall make use in a quantitative way of this property by taking 
advantage of 
one of the remarkable relations due to the ``holographic renormalization''. 
Using the 
Fefferman-Graham coordinate system for the metric
$$
{
ds^2=\f{{g_{\mu\nu}(z)}\ dx^\mu dx^\nu+dz^2}{z^2}}\ . $$
One can write
\beg
\ {g_{\mu\nu}^{}=g^{(0)}_{\mu\nu} 
{(=\eta_{\mu\nu})}+z^2_{} 
g^{(2)}_{\mu\nu} 
{(=0)}+z^4 
\cor{T_{\mu\nu}} + 
z^6 g^{(3)}_{\mu\nu} {\ldots} + }\ ,
\label{holography}
\ee
where $g_{\mu\nu}$ is the bulk metric in 5 dimensions, $\eta_{\mu\nu},$ the 
boundary 
metric in physical (3+1)  Minkowski space and $\cor{T_{\mu\nu}},$ the v.e.v. of 
the 
physical energy-momentum tensor. The higher coefficients of the expansion over 
the 
fifth dimension $z$ can be obtained by the Einstein equations in the bulk 
provided 
the 
boundary  energy-momentum tensor fulfils the  zero-trace and continuity 
equations. It is important to note that the relation (\ref{holography}) to be 
valid requires for the boundary energy-momentum tensor, by consistency
$$\ {
{T^{\mu^{}}{_\mu}
=0^{}_{} \ \  \ \ ;\ \  \ \ {\cal D}_\nu^{} T^{\mu\nu}=0^{}
}}\ ,$$
which are nothing else	 than the properties of a physical 4-d $T^{\mu\nu}$ 
with the zero trace condition of a conformal theory, verified $e.g.$ by the 
perfect fluid.

The interesting observation  on which we shall elaborate, namely that there is a 
non-trivial dual relation between a perfect fluid at rest in (3+1) dimensions 
and  
a 
static 5d black hole in the bulk can be proven 
using  
holographic renormalization. Indeed, let us consider the perfect 
fluid 
with a stress-energy tensor equipped with diagonal elements $$
\ {{ 
\cor{T_{\mu\nu}}\propto g^{(4)}_{\mu\nu} =
\left(\begin{tabular}{cccc}
$3/z_0^4=\epsilon$ & 0  & 0 & 0\\
0 & $1/z_0^4=p_{1}$  & 0 & 0\\
0 & 0  & $1/z_0^4=p_{2}$ & 0\\
0 & 0  & 0 & $1/z_0^4=p_{3}$\\
\end{tabular}\right)\ ,
}}
$$ where $\epsilon$ is the energy density and $p_1=p_2=p_3= p$ is the pressure 
density.
One can resum 
 the whole
holographic 
expansion (\ref{holography}) and get  the following bulk 
metric in Fefferman-Graham coordinates
\beg
\label{e.bhfef}
ds^2=-\f{(1-z^4/z_0^4)^2}{(1+z^4/z_0^4)z^2}\ dt^2
+(1+z^4/z_0^4)\f{dx^2}{z^2}+ \f{dz^2}{z^2}\ .
\ee

\vspace{.1cm} 
{\bf [Exercise 12.1: Recover the Energy-Momentum tensor corresponding to the 
metric (\ref{e.bhfef}), by using the expansion (\ref{holography})]}
\vspace{.1cm} 

A change of variable  $z\!\to\!\zt\equiv  
{z}/{\scriptstyle{\sqrt{1+\f{z^4}{z_0^4}}}}$ gives
\beg 
ds^2=-\f{1-\zt^4/\zt_0^4}{
\zt^2} dt^2
+\f{dx^2}{\zt^2}+\f{1}{1-\zt^4/\zt_0^4
} \f{d\zt^2}{\zt^2}\ ,
\label{metric}
\ee 
where one recognizes the Black Hole, in fact a black brane, with a static  
horizon 
at 
$\zt_0$ in 
the 5th dimension.

\vspace{.1cm}
{\bf [Exercise 12.2: prove the equivalence of the metric (\ref{e.bhfef}) and 
(\ref{metric}) by the change of variable $z\!\to\!\zt$]}
\vspace{.1cm}
 
In fact there exists a one-to-one correspondence between the 
thermodynamic properties of the  Black Hole ($BH$) and those of the perfect 
fluid 
($PF$), 
namely its {temperature} ($ T_{BH} =  
 \epsilon^{\f{1}{4}} = T_{PF}$) and {entropy} ($  S_{BH} \sim Area 
= \epsilon^{\f{3}{4}}=S_{PF}$).

It is in this context of a static Black Hole configuration that one can go 
further 
than 
the perfect fluid approximation and derive the viscosity 
using 
the Kubo formula. Indeed, the duality properties extend to a relation between 
the 
correlators of the energy-momentum tensor in two  space-time points  at zero 
frequency 
$\om=0$ and the absorption cross section $\sigma_{abs}$ of a graviton by the 
static BH 
in 
the 
bulk. One writes
\beg  \sigma_{abs}(\om)\propto \int d^4x\ \f 
{e^{i\om t}}{\om} \ 
{\cor{\left[T_{x_2x_3}(x),T_{x_2
x_3}(0)\right]}}
\Rightarrow 
\f{\eta}{S}\equiv \f 
{\sigma_{abs}(0)/16 \pi G}
{A/4 G}=\f 1{4\pi}\ ,
\label{viscosite}
\ee 
where $S=S_{BH}\equiv A/4 G$ is the famous entropy-area relation of a Black 
Hole. 
From this 
relation, and putting numbers, it appears that the viscosity is weak, much 
weaker 
than 
the one computed in the weak coupling regime and eventually realizing an 
absolute 
viscosity 
lower 
bound.

\vspace{.1cm} 
\vspace{.1cm}  {\bf \numero{12}}{\it QGP and Black Holes: From Statics to 
Dynamics}\vspace{.1cm}\vspace{.1cm}

\begin{figure}[ht]
\begin{center}
\includegraphics[width=18pc]{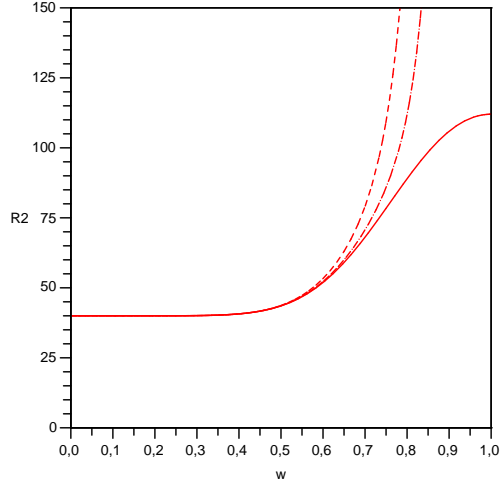}
\caption{{\it The curvature scalar  $\rsq$}. The singular structure of the 
Riemann 
scalar at the horizon apart from the perfect fluid case is   exemplified  for $s 
= \f 
43 {\pm .1}\ .$ hence,
Nonsingular 
Geometry (absence of naked singularity , $i.e.$ not hidden within the BH 
horizon) implies the Perfect Fluid condition in the considered family of 
behaviours at 
large proper time.}
\label{ricci}
\end{center} 
\end{figure}
The previous results were obtained for static configurations, $i.e.$ for a 
thermalized QGP at rest. In order to take into account, as much as possible, the 
actual 
kinematics 
of a heavy-ion collision, it is required to introduce the proper time expansion 
of 
the 
plasma. On the gravity side, it calls for studying non-equilibrium geometries, 
eventually 
of 5d BH configurations, which represent in itself a non-trivial and 
interesting 
issue.
Dual geometries to the 
standard 
``Bjorken flow'' where recently constructed. The Bjorken flow  is the 
description of a boost-invariant 
expansion 
of the QGP, which is expected to correspond to the physical situation in the 
central 
rapidity region of the collision. In this context the questions why the QGP 
fluid 
appears 
to be nearly perfect (small viscosity) and why its thermalization time can be 
short  have been adressed. 

Let us consider the equations obeyed by a physical energy-momentum 
tensor expressed in the $\left\{ \tau,\eta,x\!=\!x_1\!=\!x_2\right\}$ coordinate 
system:
\beg
\label{e.tgen1}
\ {
{\begin{tabular}{c}
$T^{\mu}_{\mu}\equiv -T_{\tau\tau}+\f{1}{\tau^2} T_{\eta\eta}+{2}T_{xx}=0$ \\ \\
${\cal D}_\nu T^{\mu\nu}\equiv \tau \f{d}{d\tau} T_{\tau\tau} 
+T_{\tau\tau}+\f{1}{\tau^2} T_{\eta\eta}=0$ \\
\end{tabular}}
}
\ee
In a boost-invariant framework, one may consider a general family of solutions 
of proper time dependent, boundary energy-momentum 
tensors
\beg\label{e.tgen}
 T_{\mu\nu}\! = \!
{\left(\begin{tabular}{cccc}
$f(\tau)$ & 0 & 0 & 0 \\
0 & $-\tau^3 \f{d}{d\tau} f(\tau)\!-\!\tau^2 f(\tau)$ & 0 & 0 \\
0 & 0 &  $f(\tau)\!+\! \f{1}{2}\tau \f{d}{d\tau} f(\tau)$  & 0 \\
0 & 0 &  0  & ... \\
\end{tabular}\right)}
\ee
where the function 
 $ f(\tau) \propto \tau^{-s},$ satisfying the positivity condition 
$${T_{\mu\nu}t^\mu 
t^\nu 
 \geq 0 \Rightarrow 0<s<4
}\ , $$ corresponds to an interpolation 
between 
different relevant regimes, namely

\vspace{.2cm}
 $ f(\tau) \propto \tau^{-\f 43}:$\ \ Perfect fluid\ \ \ \ \ \  
$\epsilon=p_1=p_2=p_3$\  \hfill

$ f(\tau) \propto 
\tau^{-1}:$ \ \ {Free streaming} \ \ \ $\epsilon=p_2=p_3;\ p_1=0$\ \hfill 

$ f(\tau) \propto 
\tau^{-0}:$ \  {``Full anisotropy'' }$\epsilon=p_\bot=-p_{L}$\hfill

\vspace{.2cm} 

Using the holographic renormalization to compute the coefficients of  the 
corresponding 
metrics in the expansion on the fifth dimension and after resummation, it was 
possible to 
solve the dual geometry for given  $s$ at asymptotic  proper time $\tau.$ It 
reveals 
the 
existence of a scaling property of the solutions in terms of  the proper time 
dependent variable  
\vspace{-.2cm} 
$$
{ v=\f{z}{\tau^{1/3}}}\ .
$$

Analyzing the family of solutions as a function of $s,$ it appears that the only 
non-singular solution for invariant scalar quantities (here the square of the 
Ricci 
tensor
$\rsq=R^{\mu\nu\alpha\beta}R_{\mu\nu\alpha\beta},$ see Fig.~\ref {ricci}), is 
obtained for $s=4/3.$ Indeed, we find in Fefferman-Graham coordinates:
$$
\ {{ds^2=\f{1}{z^2} 
\left[- \f{\left( 1\!-\!\f{e_0}{3}\ 
\f{z^4}{\tau^{4/3}}\right)^2}{1\!+\!\f{e_0}{3}\f{z^4}{\tau^{4/3}} } 
d\tau^2\!+\!
\left( \textstyle 1\!+\!\f{e_0}{3} \f{z^4}{\tau^{4/3}}\right) (\tau^2 d\eta^2 
\!+\!dx^2) 
\right] \!+\! \f{dz^2}{z^2}
}}
$$
which is similar to the metrics of the static Black Hole (\ref{e.bhfef}), but 
substituting $z_0\to {z^4}/{\tau^{1/3}}.$
This solution is the only one of the family corresponding  
to 
a Black Hole 
moving 
away in 
the fifth dimension. Hence 
the perfect-fluid case is 
singled out  and the  moving BH in the bulk corresponds through duality  
to the expansion of the QGP taking place in the 
boundary. 
Consequently, the BH horizon moves as $z_h(\tau) \propto
 \tau^{\f{1}{3}},$
 the temperature as $T(\tau) \sim  
{1}/{z_h} \sim \tau^{-\f{1}{3}},$ and the entropy stays constant since $ S(\tau) 
\sim 
Area 
\sim \tau \cdot {1}/{z_h^3} \sim 
const.$ Note that again the physical thermodynamical variables of the QGP are 
the 
same as 
those one may attribute to the BH in the bulk (with the reservation that 
thermodynamics of a moving BH may rise   non-trivial interpretation problems). 
Hence 
one finds a 
concrete 
realization of the idea of a duality between the QGP 
formation and a 
moving Black Hole.

\vspace{.1cm}
\vspace{.1cm}  {\bf \numero{13}}{\it Thermalization and 
Isotropization}\vspace{.1cm}\vspace{.1cm}

There has been a lot of activity along the lines of the AdS/CFT correspondence 
and 
its 
extensions to various geometric configurations. Dual studies of jet quenching, 
quark
dragging , etc... have been and are still being performed. 
Sticking 
to the configurations 
corresponding to  an expanding plasma and going beyond the first order terms in 
proper time, one has obtained results on the viscosity, 
confirming the 
universal value (\ref{viscosite}), on the relaxation time of the 
plasma and 
very recently on the inclusion of flavor degrees of freedom.

Let us finally focus on the thermalization problem, which can be usefully taken 
up 
using 
the Gauge/Gravity duality in the strong coupling hypothesis. the problem is to 
give 
an 
explanation to the strikingly small thermalization time required for the 
formation 
of a 
QGP as can be abstracted from the experimental observations. Analyzing the 
stability 
of the expanding plasma configuration, it has been found that performing a small 
deviation 
from the 
BH metric by coupling with a scalar field and analyzing  the 
corresponding 
{\it quasi-normal 
modes} defining the way how the system relaxes towards its initial state, one 
finds a 
numerically small value of the relaxation time in units of the local 
temperature. Even if a definite value of this  relaxation time cannot be 
inferred 
at this stage due to  scale-invariance, this result was suggestive of a 
stability  
of the QGP  in the strong coupling regime
with 
respect to perturbations out of equilibrium.

In order to go further, one has to deal with the problem of the QGP evolution at 
small 
proper times. The holographic renormalization 
program has 
been pursued for the small proper time expansion. Relaxing the selection of the 
appropriate 
metric by requiring  only the metric tensor to be a real and single valued 
function 
of 
the 
coordinates everywhere in the bulk, one finds an unique solution corresponding 
to 
the 
``fully anisotropic case'' $s=0.$ 
\begin{figure}[ht]
\begin{center}
\includegraphics[width=18
pc]{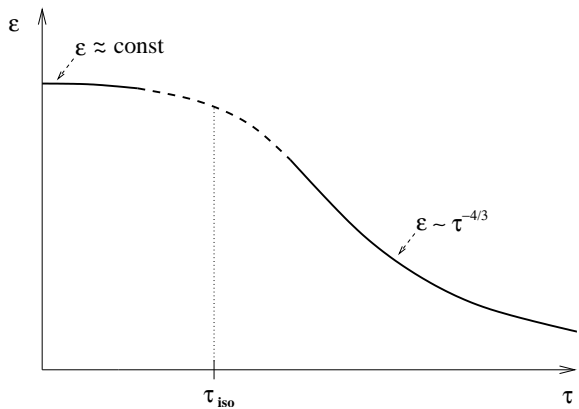}
\end{center}
\caption{{\it Evaluation of  the isotropization/thermalization time}. The 
behaviour at 
large proper times and the one found at small ones are matched with the 
condition that 
the branching points where the solutions become multi-valued are avoided. the 
matching 
happens for a value $\tau_{iso}$ whose range gives an evaluation of the 
isotropization 
time.}\label{therma}
\end{figure}
In the same paper, an evaluation of the range of the isotropization time has 
been 
proposed, by extrapolation of realistic estimates abstracted from experiments to 
the supersymmetric case. The 
idea is to match the large and small proper time regimes at some value of the 
proper time  $\tau_{iso}.$ This proper time is mathematically defined as the 
crossing value for the branch-point singularities of both regimes. Physically, 
it 
is expected to give an estimate of the proper time range during which which the
medium evolves from the full anisotropic regime (small $\tau$) to the perfect 
fluid 
one (large $\tau$).

In order to give an idea of the possible physical implications of this strong 
coupling scheme, let us shortly reproduce the argument.  Implementing the 
estimated 
physical value of the 
energy density at some proper time
($e.g.\  \epsilon(\tau)\  = \ e_0\ 
\tau^{4/3}\vert_{\tau=.6} 
\sim 15\ {\scriptstyle GeV fermi^{-3}}$) one finds 
\beg
\tau_{iso}=\left(\f {3 
N_c^2}{2\pi^2e_0}\right)^{3/8}\sim \ .3\  
{\scriptstyle fermi}\ .
\label{iso}
\ee
 This short isotropization time thus seems a 
characteristic feature of the strong coupling scenarios. It is clear that more 
realistic estimates should take into account less idealized dual models, 
corresponding  to  QCD, such as the lack of supersymetry and 
the finite numbers of colors. However, the non confined character of the QGP and 
the robustness of some predictions (such as the $\eta/S$ ratio) may give some 
confidence that this short isotropization time could be a reasonable estimate at 
strong coupling. 
 
\vspace{.1cm}
\vspace{.1cm}  {\bf \numero{14}}{\it Outlook}
\vspace{.1cm}\vspace{.1cm}

From the present rapid (and partial) survey of some of the results obtained in 
the 
AdS/CFT approach to the formation and expansion of the Quark-Gluon plasma in 
heavy-ion collisions, it appears that the Gauge Gravity correspondence is a 
promising way to explore some features of QCD at strong coupling. Indeed some 
general features of  this correspondence, relating at long distances the closed 
and open string geometries (see Fig.~\ref{opclo1}) are expected to be valid in 
principle 
for various dual schemes  and thus, hopefully, QCD.

In practice, the quantitative dual schemes have been more precisely elaborated 
for 
the specific AdS/CFT case,  $i.e.$ the gauge theory with ${\cal N}=4$ 
supersymmetries. Among the results, it gives a 
calculable link between the hydrodynamic  quasi-perfect fluid behaviour on the 
``gauge theory side'' with a BH geometry in the higher dimensional ``gravity 
side'' 
in an AdS background. This relation can be extended from the static case to a 
dynamical regime reflecting (within the AdS/CFT framework) the relativistic 
expansion of the corresponding quark-gluon plasma. This, and many other 
applications, some of them using more complex geometries, less supersymmetric 
backgrounds and examining other observables, gives hope for the fruitful  
possibilities of the Gauge/Gravity approach to the QGP formation.

As an outlook, it is worth mentionning some of the possible new directions of 
study
one is led to consider. Starting with the more technical ones,  it is known that 
the Bjorken flow is not exactly verified in heavy-ion collisions, since the 
observed 
distribution of particles is nearly gaussian  in rapidity and thus not 
reflecting 
exactly the  boost-invariance of the Bjorken flow. It would be interesting to 
investigate dual properties for non-boost invariant flows, such as the {\it 
Landau 
flow}. On a more general  ground, the whole approach 
still 
concerns only the hydrodynamical stage of the QGP expansion. It would be 
important 
to attack both the initial (partonic) and final (hadronic) stages of the 
reaction 
in the same framework and thus the problem of {\it phase transitions} during the 
collision. Finally, one would like to have more realistic dual frameworks 
including 
a finite number of colors, flavor degrees of freedom and no (or broken) 
supersymmetry. 

\section*{Acknowledgements}
I would like to thank Romuald Janik (Jagelonian University, Cracow) for a 
fruitful and inspiring long-term collaboration between a string theorist and a 
particle phenomenologist. Emmanouil Saridakis, Alessandro Papa and Christophe 
Royon are thanked for helping to correct the manuscript.

\vfill
\eject
\vspace{2cm}

\BL \centerline{\bf  {Note on Lecture I}}\EL 

\vspace{1cm}


\vspace{1cm}

The lecture I is based on an introduction to string theory $via$ the strong 
interaction peoblems, and thus follows more or less the historical development. 
Starting with  Veneziano's initial paper, 
 we were largely inspired by the two next references  of the 
list. The first one is a book which has the merit to give the whole derivation 
(with  detailed calculations) of the  introduction to string theory $via$ the 
strong interaction paradigm. The second one is very helpful in understanding the 
story of the discovery of string theory and containing an abundant  and 
documented list of references placed in their historical perspective (as 
captivating as a detective novel). Moreover a large part of the derivation that 
I used is here given with useful details.

Nowadays, string theory has developped as an autonomous theory, and many deep 
properties and its modern ways of  introduction are far from the initial 
problematics. For primary instance the intimate connection of string 
theory to 2-dimensional conformal field theory do (and did) not appear 
immediately from the construction of the Hilbert space through scattering 
amplitudes as sketched in lecture I.  Hence it is strongly advised for the 
students who would like to learn string theory to refer to modern textbooks. My 
colleagues string theorists would, among others, recommend the two last 
references of the list, and find there the appropriate references. I make my 
apologizes  to all contributors to string theory for not quoting them {\it 
directly} in the lecture for sake of simplicity (and possible misunderstanding 
from my part).

\vspace{2cm}
\bc \BL {\bf  {Note on Lecture II}}\EL \ec

\vspace{2cm}
The lecture II is based on an elementary introduction to the AdS/CFT 
correspondence. 
In a considerably shortened and subjective selection on the enormous literature 
on the 
subject the three first references are the original ones. The fourth one is a 
now 
rather old but still precious and readable review on the subject. In the general 
argumentation about gauge/gravity duality, we were inspired by the nice  
introduction 
of Ref.[10]. 

In introducing some concepts useful in the approach to the formulation of 
hadronic 
scattering amplitudes at strong coupling, we used mainly the duality vocabulary 
related 
to Wilson loops. We quote some relevant original  references, and for the rest 
refer to   the 
review [7].

The specific approach to dipole scattering amplitudes described in the  text 
comes 
from a collaboration with Romuald Janik, as quoted in the references.

\vspace{2cm}
\bc \BL {\bf  {Note on Lecture III}}\EL \ec
\vspace{1cm}

The lecture III is based on a series of papers using the AdS/CFT correspondence 
as a 
laboratory for the flow of a strongly coupled quark-gluon plasma. It relies both 
on 
the phenomenological validity of the relativistic hydrodynamic approach to the 
real 
experiments and its surprisingly good  relevance for the 
Gauge/Gravity 
correspondence.

The first set of references recall some basic facts about 1+1 relativistic 
hydrodynamic equations applied to particle collisions concerning the canonical 
Bjorken 
and Landau flows and including a recent interpolating family of exact solutions.

The second set of references begins with  holographic renormalization, which is 
a 
basic tool used in holographic reconstruction followed by original references on 
the 
emergence and calculations of the dual Black Hole geometry of a static plasma.
In a third set are given some the original references corresponding to the 
expanding 
plasma geometry and its evolving black hole dual.

We display at the end a very limited number of references concerning  other 
interesting aspects of the quark gluon plasma problems treated from the point of 
view 
of the AdS/CFT correspondence and which were not addressed in this lecture. 
However 
the general set-up described here may help for the understanding of the various 
approaches.

\eject

\end{document}